\documentclass[floatfix,amsmath,amssymb,superscriptaddress,aip,jcp,reprint,widetext,longbibliography]{revtex4-2}

\usepackage[utf8]{inputenc}
\usepackage{graphicx}
\usepackage{xcolor}
\usepackage{notes2bib}
\usepackage{bbold}
\usepackage{mathtools}
\usepackage{scalerel}
\usepackage{bm}
\newcommand{\smallx}{\scaleobj{.65}{{\mathcal{X}}}}

\begin{document}

\title{Nonlinear Langevin functionals for a driven probe}

\author{Juliana Caspers}
\email{j.caspers@theorie.physik.uni-goettingen.de}
\affiliation{Institute for Theoretical Physics, Georg-August-Universit\"{a}t G\"{o}ttingen, 37073 G\"{o}ttingen, Germany}

\author{Matthias Krüger}
\email{matthias.kruger@uni-goettingen.de}
\affiliation{Institute for Theoretical Physics, Georg-August-Universit\"{a}t G\"{o}ttingen, 37073 G\"{o}ttingen, Germany}

\begin{abstract}
    When a probe particle immersed in a fluid with nonlinear interactions is subject to strong  driving, the cumulants of the stochastic force acting on the probe are nonlinear functionals of the driving protocol.
    We present a Volterra series for these  nonlinear functionals, by applying nonlinear response theory in a path integral formalism, where the emerging kernels are shown to be expressed in terms of connected equilibrium correlation functions. The first cumulant is the mean force, the second cumulant characterizes the non-equilibrium force fluctuations (noise), and higher order cumulants quantify non-Gaussian fluctuations. We discuss the interpretation of this formalism in relation to Langevin dynamics. We highlight two example scenarios of this formalism: i) For a particle driven with prescribed trajectory, the formalism yields the non-equilibrium statistics of the interaction force with the fluid. ii) For a  particle confined in a moving trapping potential, the formalism yields the non-equilibrium statistics of the trapping force. 
    In simulations of a model of nonlinearly interacting Brownian particles, we find that nonlinear phenomena, such as shear-thinning and oscillating noise covariance, appear  in third or second order response, respectively.
\end{abstract}

\maketitle

\section{Introduction}

The jittery motion of a Brownian particle in a Newtonian fluid is well described by a  Markovian Langevin equation, since the time scales of the relaxation of the solvent and that of a Brownian particle are separated~\cite{dhont_introduction_1996}. 
More complex fluids, such as colloidal suspensions, polymer solutions or others, can, in contrast, exhibit structural relaxation times on the order of seconds or longer~\cite{dhont_introduction_1996,baiesi_rise_2021,ginot_recoil_2022}. 
 Brownian particles in such surroundings thus experience memory, and a non-Markovian description is required, achieved, e.g., via projection operator techniques~\cite{zwanzig_memory_1961,mori_transport_1965,zwanzig_nonequilibrium_2001,grabert_projection_1982}, in which a  set of degrees of freedom are projected out.
For linear systems, such as the famous Caldeira-Leggett model~\cite{caldeira_influence_1981,zwanzig_nonequilibrium_2001}, using a bath of harmonic oscillators,  such techniques can be applied explicitly to obtain  exact non-Markovian Langevin equations.
Projection-operator techniques have also been applied in mode-coupling theory (MCT)~\cite{gotze_complex_2008,janssen_mode-coupling_2018}, including predictions of the sub-diffusive plateau in mean squared displacements~\cite{fuchs_asymptotic_1998} observed experimentally~\cite{van_zanten_brownian_2000,lu_probe_2002,van_der_gucht_brownian_2003,caspers_how_2023}.

Going beyond equilibrium and harmonic cases, i.e., by considering a driven probe particle surrounded by fluid particles with anharmonic interactions, the probe will excite the bath (the fluid particles) out of equilibrium, such that the probe may in general be expected to be subject to non-equilibrium, non-Markovian, and non-Gaussian forces and noise.
This can manifest  itself, for example, in particle oscillations~\cite{doerries_correlation_2021,venturelli_memory-induced_2023}, shear-thinning~\cite{squires_simple_2005,gazuz_active_2009,harrer_force-induced_2012} observed in experiments with micellar fluids~\cite{jayaraman_oscillations_2003,handzy_oscillatory_2004,berner_oscillating_2018,jain_two_2021}, superdiffusive behavior~\cite{harrer_force-induced_2012,winter_active_2012,benichou_geometry-induced_2013} or effective temperature \cite{wilson_small-world_2011,demery_driven_2019,cugliandolo_effective_2011,puglisi_temperature_2017}.

Theoretical description of nonlinear stochastic dynamics of a driven probe is challenging, both conceptually as well as practically, and subject of intense research in recent years.  The mentioned projection operator techniques have been extended to nonlinear and non-equilibrium cases~\cite{brader_first-principles_2008, kruger_tagged-particle_2011, gazuz_active_2009,gazuz_nonlinear_2013,gruber_active_2016,meyer_non-stationary_2017,te_vrugt_mori-zwanzig_2019,meyer_dynamics_2019,glatzel_interplay_2021,vroylandt_position-dependent_2022,schilling_coarse-grained_2022,netz_derivation_2023};
field theory approaches~\cite{demery_driven_2019,venturelli_memory-induced_2023}, dynamical density functional theory \cite{penna_dynamic_2003, rauscher_dynamic_2007,gutsche_colloids_2008,de_las_heras_velocity_2018,de_las_heras_flow_2020,schmidt_power_2022,schilling_coarse-grained_2022}, lattice models~\cite{benichou_geometry-induced_2013} and nonlinear response theory~\cite{asheichyk_brownian_2021,kruger_modified_2016,muller_brownian_2020} within  path integral formalism~\cite{colangeli_meaningful_2011,basu_frenetic_2015,maes_response_2020} have been applied.

In this manuscript, we present a formalism, based on the methods developed in Refs.~\onlinecite{kruger_modified_2016,muller_brownian_2020},  for systematic expansion of the force cumulants experienced by a probe driven in a fluid of nonlinearly interacting Brownian particles.
We therefore derive Volterra series for these cumulants, and find explicit expressions for the respective kernels. Notably, it is the connected equilibrium correlation functions that appear in these kernels. We interpret the resulting Langevin functionals in relation to Langevin dynamics.
We evaluate the kernels in a specific nonlinear  model, the stochastic Prandtl-Tomlinson model~\cite{muller_properties_2020,jain_two_2021,jain_micro-rheology_2021}, and discuss two different driving modes. 

This paper is organized as follows: we start with an explanation of the setup in Sec.~\ref{chap:Setup}. 
In Sec.~\ref{chap:Derivation} we move to a path integral representation and present the expansion scheme.
We derive the Volterra series for the nonlinear Langevin functionals, the interpretation of which we discuss in Sec.~\ref{chap:Embedding}.
In Sec.~\ref{chap:SteadyShearSPT} we discuss the application to a specific system, the stochastic Prandtl-Tomlinson model and confirm theoretical predictions with simulations of steady velocity driving in two scenarios of direct and indirect driving.

\section{Setup}\label{chap:Setup} 

\begin{figure}
    \centering
    \includegraphics{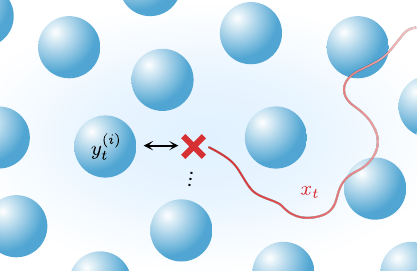}
    \caption{Setup of $N$ Brownian overdamped degrees of freedom (blue) and one external control parameter $x_t$ (red). 
    The $i$-th Brownian degree of freedom $y^{(i)}$ experiences a potential gradient $-\partial_{y^{(i)}}U(\{y^{(i)}-x\})$ that can encompass pairwise interactions and coupling to the control parameter. }
    \label{fig:MasterSetup}
\end{figure}

We consider $N$ Brownian, overdamped degrees of freedom $\textbf{y}$, with $y^{(i)}$, $i=1\dots N$, the $i$th component, see Fig.~\ref{fig:MasterSetup}. These are subject to a potential $U(\{y^{(i)}_t-x_t\}) $, which encompasses  mutual interactions between the Brownian degrees as well as interactions with an external control parameter $x_t$, with  time index $t$: $x_t$ is deterministic, and it takes the role of the time-dependent perturbation that is applied to the system (the protocol). Degree $i$ is subject to a bare friction coefficient $\gamma_i$, so that it follows the equations of motion,
\begin{align}
    \gamma_i \dot{y}_t^{(i)} &= -\partial_{y^{(i)}_t}U(\{y^{(i)}_t-x_t\}) +\xi_t^{(i)} \label{eq:MasterLangevin}
\end{align}
with Gaussian white noise $\xi_t^{(i)}$,
\begin{align}
    \left\langle \xi_t^{(i)}\right\rangle=0 \quad \text{and} \quad \left\langle \xi_t^{(i)}\xi_{t'}^{(j)}\right\rangle = 2  k_B T  \gamma_i \delta_{ij}\delta(t-t').
\end{align}
The observable of our interest is $\partial_{x}U(\{y^{(i)}-x\})$, i.e., (minus) the stochastic force that acts on the controlled parameter $x$ \footnote{The minus sign is included because it will yield the familiar sign convention for the case of indirect driving introduced below.}. We will find a Volterra series for its cumulants. This setup encompasses, among others, the following  two explicit cases. 
In example (i) $x$ parametrises the position of a trapping potential that acts on one of the particles, say particle 1~\cite{wilson_small-world_2011, berner_oscillating_2018,jain_two_2021}. In this case, $\partial_{x}U(\{y^{(i)}-x\})$ is the force acting on this particle by the trapping potential, accessible in experiments~\cite{wilson_small-world_2011,berner_oscillating_2018,jain_two_2021}.  
In example (ii), $x$ is the position of a probe particle that is driven with controlled velocity. In this case, $\partial_x U(\{y^{(i)}-x\})$ is (minus) the stochastic force the driven particle feels from the Brownian degrees of freedom, i.e., the resulting ''frictional force``.
This case may be accessible in experiments with feedback control \cite{CBprivate}.

We note that the observable of interest, $\partial_{x_t}U(\{y^{(i)}_t-x_t\})$, depends explicitly on time, via $x_t$. This makes a perturbative expansion of it in powers of displacements of $x_t$  cumbersome \cite{basu_frenetic_2015,maes_response_2020,muller_brownian_2020}, because the expansion also includes explicit expansion of the observable. Identifying entropy production (see below) appears also nontrivial in this case \cite{maes_response_2020,maes_frenesy_2020}. It is thus beneficial to introduce relative coordinates $q^{(i)}=y^{(i)}-x$ and to transform to a frame comoving with $x_t$. In this frame, 
the $q^{(i)}$ follow
\begin{align}
    \gamma_i \dot{q}_t^{(i)} &= -\gamma_i \dot x_t-\partial_{q^{(i)}_t}U(\mathbf{q}_t) +\xi_t^{(i)}. \label{eq:RelativeMasterLangevin}
\end{align}
We note that the simple transformation of Eq.~\eqref{eq:MasterLangevin} to the moving frame is the main reason to restrict to overdamped, Brownian degrees of freedom. 

Using translational invariance of the potential $U$, the observable  $\partial_{x_t} U(\{y_t^{(i)}-x_t\})$ reads in these coordinates 
\begin{align}
\begin{split}
F_t&\equiv\partial_{x_t} U(\{y_t^{(i)}-x_t\})=-\sum_i \partial_{y^{(i)}_t} U(\{y^{(i)}_t-x_t\})\\&=
-\sum_i \partial_{q^{(i)}_t} U(\mathbf{q}_t).
\end{split}
\label{eq:TranslationInvariancePotential}
\end{align}
The explicit dependence on time $t$ thus disappeared in this frame, easing the expansion.

\section{Volterra Series}\label{chap:Derivation}

We aim to expand the cumulants of the force $\partial_{x_t}U(\{y_t^{(i)}-x_t\})=F(\textbf{q}_t)\equiv F_t$ in a Volterra series in the driving by applying a path integral formalism~\cite{basu_frenetic_2015,maes_response_2020}. 
Therefore consider paths $\omega = (\textbf{q}_s, t_0 \leq s \leq t)$ of the $N$ Brownian degrees of freedom over a time-interval $[t_0,t]$ with
$\textbf{q}_s$ the state of $N$ Brownian degrees of freedom at time $s$, prepared, at $t_0$, in the equilibrium distribution corresponding to $x$ at rest at $x_{t_0}$.  
The expectation value of observable $O(\omega)$ under the protocol $\smallx = (x_s,t_0\leq s \leq t)$ is given by the path integral
\begin{align}
    \langle O(\omega)\rangle = \int \mathcal{D}\omega\, \mathcal{P}(\smallx,\omega) O(\omega),
    \label{eq:pathintegral}
\end{align}
with $\mathcal{D}\omega$ the path element and
$\mathcal{P}(\smallx,\omega)$ the path weight, a functional of $\omega$ depending on protocol $\smallx$.
The latter is expressed in terms of the equilibrium path weight $\mathcal{P}_\mathrm{eq}$ and perturbation action $\mathcal{A}$,
\begin{align}
    \mathcal{P}(\smallx,\omega) = e^{-\mathcal{A} (\smallx,\omega)} \mathcal{P}_\mathrm{eq}(\omega).
    \label{eq:Pathweight}
\end{align}
$\mathcal{P}_\mathrm{eq}(\omega)$ contains the Boltzmann distribution. The action $\mathcal{A}$ follows from the difference of the Onsager-Machlup action functional~\cite{onsager_fluctuations_1953} corresponding to Eq.~\eqref{eq:RelativeMasterLangevin} for finite and zero velocity of $x$, 
\begin{align}
\begin{split}
    \mathcal{A}(\smallx,\omega) = \sum_i \frac{\gamma_i}{4 k_B T} \int_{t_0}^t \mathrm{d}s\, \Bigg[-\left(\dot{q}_s^{(i)} + \frac{\partial_{q^{(i)}_s}U(\mathbf{q}_s)}{\gamma_i}\right)^2 &\\
   +\left(\dot{q}_s^{(i)} + \dot{x}_s  + \frac{\partial_{q^{(i)}_s}U(\mathbf{q}_s)}{\gamma_i}\right)^2 \Bigg]&
    \label{eq:MasterOMAction}
    \end{split}\\
  = \frac{1}{2k_B T} \int_{t_0}^t \mathrm{d}s \left[-\dot{x}_s F_s+\sum_i \frac{\gamma_i}{2 } \left(2 \dot{x}_s  \dot{q}_s^{(i)}  + \dot{x}_s^2 \right) \right] &\label{eq:PerturbedAction}.
\end{align}
Eq.~\eqref{eq:PerturbedAction} contains the driving $\dot{x}$ up to second order for the model of Brownian degrees. 
It is insightful to decompose $\mathcal{A}$ into its time-antisymmetric part $S(\omega)$, termed the \textit{entropic} component, and its time-symmetric part $D(\omega)$. The latter, named frenetic part~\cite{baiesi_fluctuations_2009}, has no thermodynamic meaning, and we refer to it as \textit{non-thermodynamic} (NT) component in the following: $\mathcal{A} \coloneqq D-S/2$~\cite{maes_time-reversal_2003,colangeli_meaningful_2011,baiesi_fluctuations_2009,basu_frenetic_2015,maes_response_2020,maes_frenesy_2020,muller_brownian_2020}.
$S(\theta \omega) = -S(\omega)$ and $D(\theta \omega) = D(\omega)$, with time reversal operator $\theta$, i.e.,  $(\theta \smallx,\theta \omega)_s = \pi (x_{t-s},\mathbf{q}_{t-s})$
with $\pi$ denoting the kinematical time-reversal. 
Taking into account the change of sign of velocities under time reversal, we can read off the respective entropic and NT components (denoting by $'$ and $''$ the terms in first and second order of driving, respectively),
\begin{align}
   S= S' &= \frac{1}{k_B T} \int_{t_0}^t \mathrm{d}s\, \dot{x}_s F_s\label{eq:SMasterLangevin}\\
    D' &= \sum_i \frac{\gamma_i}{2 k_B T} \int_{t_0}^t \mathrm{d}s\, \dot{x}_s \dot{q}_s^{(i)} \label{eq:D1MasterLangevin}\\
    D'' &= \frac{\sum_i \gamma_i}{2 k_B T} \int_{t_0}^t \mathrm{d}s \int_{t_0}^t \mathrm{d}s' \, \dot{x}_s \dot{x}_{s'} \delta(s-s') \label{eq:D2MasterLangevin}.
\end{align}
Eq.~\eqref{eq:SMasterLangevin} thus corresponds to the entropy change with respect to the reference equilibrium ensemble, $S = \ln \frac{\mathcal{P}(\smallx,\omega)}{\mathcal{P}(\theta\smallx,\theta\omega)}$~\cite{maes_response_2020,seifert_stochastic_2012}. 
It is inherently linear in the perturbation, $S = S'$, which also appears to be specific for the Brownian system~\cite{maes_response_2020}.
For the NT components, Eqs.~\eqref{eq:D1MasterLangevin} and \eqref{eq:D2MasterLangevin}, 
we introduce abbreviations $\mathcal{D}_s \coloneqq \sum_i \frac{\gamma_i}{2 k_B T} \dot{q}_s^{(i)}$ and $\mathcal{D}_{s,s'} \coloneqq \frac{\sum_i \gamma_i}{2 k_B T}\delta(s-s')$, i.e.,
\begin{align}
D' &= \int_{t_0}^t \mathrm{d}s\, \dot{x}_s \mathcal{D}_s\\ D'' &= \int_{t_0}^t \mathrm{d}s \int_{t_0}^t \mathrm{d}s' \, \dot{x}_x \dot{x}_{s'} \mathcal{D}_{s,s'}.
\end{align}
Expanding also the exponential in Eq.~\eqref{eq:Pathweight}, the response for any path observable $O=O(\omega)$ is given by~\cite{colangeli_meaningful_2011,basu_frenetic_2015,maes_response_2020}
\begin{align}
\begin{split}
\langle O\rangle &= \langle O\rangle_\mathrm{eq} - \langle (D'-S'/2)O\rangle_\mathrm{eq} \\
&\quad+ \frac{1}{2} \langle (D'^2-D''-D'S'+S'^2/4)O\rangle_\mathrm{eq}+\dots\label{eq:ResponseO}
\end{split}
\end{align}
While Eq.~\eqref{eq:ResponseO} holds for path observables, considering state observables $O_t \coloneqq O(\mathbf{q}_t)$ allows for simplifications: taking the difference $\langle O_t \rangle - \langle O(\theta \omega)\rangle$ yields, because of initial preparation in equilibrium~\cite{basu_frenetic_2015,maes_response_2020,muller_brownian_2020}
\begin{align}
    \begin{split}
        &\langle O_t\rangle - \langle O_t\rangle_\mathrm{eq} =   \langle S'O_t\rangle_\mathrm{eq} -  \langle D' S' O_t\rangle_\mathrm{eq} \\
        &\quad + \frac{ 1}{6} \left\langle \left[ 3 D'^2 - 3 D''+ \frac{S'^2}{4} \right] S'O_t \right\rangle_\mathrm{eq}+ \dots
    \end{split}
    \label{eq:ResponseStateO}
\end{align}
See Eq.~\eqref{eq:ForceResponseInfty} in Appendix~\ref{app:Gamman}~\cite{holsten_thermodynamic_2021} for the higher order terms of Eq.~\eqref{eq:ResponseStateO}, which, for any order $n$, can be given in closed form.

Notably, the terms shown in Eq.~\eqref{eq:ResponseStateO} may individually diverge for large $t$ for specific cases, see discussion in Appendix~\ref{app:ConvergenceConnectedCorr}. 
We can however show that these divergences  analytically cancel, and that  
 Eq.~\eqref{eq:ResponseStateO} can be expressed in terms of \textit{connected correlation functions}  $\langle \cdot;\dots;\cdot\rangle$, which are mathematically equivalent to joint cumulants (see Appendix~\ref{app:connectedCorr} for the definition),
 i.e. 
\begin{align}
    \begin{split}
        \langle O_t\rangle - \langle O_t\rangle_\mathrm{eq} &=   \langle S';O_t\rangle_\mathrm{eq} -  \langle D';  S';  O_t\rangle_\mathrm{eq} \\
        &\quad + \frac{ 1}{2} \left[ \langle D';D';S';O_t\rangle_\mathrm{eq} - \langle D'';S';O_t\rangle_\mathrm{eq}\right] \\
        &\quad+ \frac{ 1}{24} \langle S';S';S';O_t\rangle_\mathrm{eq}+\dots
    \end{split}
\label{eq:ResponseStateOConnectedCorrelations}
    \end{align}
    Note that we have only shown this up to including  third order, and we do not know whether the higher order terms can be written in terms of connected correlation functions as well.
For the derivation we refer to Appendix~\ref{app:connectedCorr}. 

The notion of connected correlation functions is reminiscent of connected Feynman diagrams in field theoretic approaches~\cite{kardar_statistical_2007}. It will be worth exploring this connection in future work.
We illustrate the resulting cancellations in Appendix~\ref{app:ConvergenceConnectedCorr} for the specific example discussed in Sec.~\ref{chap:SPT_Flowcurve}.

The occurrence of connected correlations  is not restricted to state observables. For  path observables $O(\omega)$ we can equally express the response (here proven up to second order) 
\begin{align}
\begin{split}
    &\langle O\rangle -\langle O\rangle_\mathrm{eq} =    \left[ \frac{1}{2}\langle S';O\rangle_\mathrm{eq}- \langle D';O\rangle_\mathrm{eq}  \right] -\frac{ 1}{2}  \langle D'';O\rangle_\mathrm{eq} \\
    & +\frac{ 1}{2} \Big[  \langle D';D';O\rangle_\mathrm{eq} -\langle D';S';O\rangle_\mathrm{eq} + \frac{1}{4} \langle S';S';O\rangle_\mathrm{eq} \Big] +\dots
    \end{split}\label{eq:ResponseOConnectedCorrelations}
\end{align}
It is also worth noting that all response relations presented in this section are independent of the assumption of Brownian overdamped degrees of freedom.
Mathematically, they are solely based on the expansion $\mathcal{A} = D' -S'/2 + D''/2 + \dots$.
To our knowledge Eqs.~\eqref{eq:ResponseStateOConnectedCorrelations} and \eqref{eq:ResponseOConnectedCorrelations}  have not been given in literature.

In the following, we will apply Eqs.~\eqref{eq:ResponseStateOConnectedCorrelations},  \eqref{eq:ResponseOConnectedCorrelations}, and \eqref{eq:SMasterLangevin}-\eqref{eq:D2MasterLangevin} to find the cumulants of the stochastic force $F_t$.
Eq.~\eqref{eq:ResponseStateOConnectedCorrelations} is used to obtain the mean force $\langle F_t\rangle$, while higher cumulants $\langle F_t;F_{t'}\rangle$, $\langle F_t;F_{t'};F_{t''}\rangle, \dots$ follow from Eq.~\eqref{eq:ResponseOConnectedCorrelations}.
We formally take the limit $t_0 \to -\infty$ in Eqs.~\eqref{eq:SMasterLangevin}-\eqref{eq:D2MasterLangevin} to allow for arbitrary driving protocols $\dot{x}_t$.
For example, steady-state averages follow then with a time independent driving velocity, and transient ones are obtained by use of a step function $\dot{x}_t \propto \Theta(t)$. 
Noting that the mean force vanishes for the equilibrium case, the resulting Volterra series take on the form (with $\beta = (k_B T)^{-1}$) 
\begin{widetext}
    \begin{align}
       \beta \langle F_t\rangle &=  \int_{-\infty}^t \mathrm{d}s\, \dot{x}_{s} \Gamma_{t-s}^{(0,1)}  +  \int_{-\infty}^t \mathrm{d}s\int_{-\infty}^t\mathrm{d}s'\, \dot{x}_{s}\dot{x}_{s'} \Gamma_{t-s,t-s'}^{(0,2)}+  \int_{-\infty}^t \mathrm{d}s\int_{-\infty}^t \mathrm{d}s' \int_{-\infty}^t \mathrm{d}s'' \, \dot{x}_{s} \dot{x}_{s'} \dot{x}_{s''} \Gamma_{t-s,t-s',t-s''}^{(0,3)} + \dots  \label{eq:MeanForce} \\
       \beta^2 \langle F_t;F_{t'}\rangle &=  \Gamma_{t-t'}^{(1,0)} + \int_{-\infty}^t \mathrm{d}s \, \dot{x}_{s} \Gamma^{(1,1)}_{t-s;t-t'} + \int_{-\infty}^t \mathrm{d}s \int_{-\infty}^t \mathrm{d}s' \, \dot{x}_{s} \dot{x}_{s'} \Gamma^{(1,2)}_{t-s,t-s';t-t'} + \dots \label{eq:ForceCov} \\
       \beta^3 \langle F_t;F_{t'};F_{t''}\rangle &= \Gamma^{(2,0)}_{t-t',t-t''} + \int_{-\infty}^t \mathrm{d}s \, \dot{x}_{s} \Gamma^{(2,1)}_{t-s;t-t',t-t''} 
       +\dots \label{eq:ForceCum3} \\
        &\;\;\vdots  \nonumber
    \end{align}
\end{widetext}
Here we have assumed the existence of response kernels $\Gamma^{(m,n)}$, with $m+1$ and $n$ indicating the order of cumulant and order of perturbation, respectively. The forms \eqref{eq:MeanForce} -  \eqref{eq:ForceCum3} and higher orders are so far pure definitions, not related to the system at hand,   
\begin{align}
\begin{split}
    \Gamma^{(m,n)}&_{t-s_1,\dots,t-s_n;t-t_1,\dots,t-t_m} \\&\coloneqq \frac{\beta^{m+1}}{n!} \frac{\delta^n}{\delta \dot{x}_{s_1}\cdots \delta \dot{x}_{s_n}} \langle F_t;F_{t_1};\dots;F_{t_m}\rangle \Big|_{\{\dot{x}\}=0}.
    \end{split}
    \label{eq:Gammamn}
\end{align}
The novelty of this manuscript is now contained in obtaining these kernels in terms of correlation functions. These follow from the above laid out expansion. Explicitly, the leading order  kernels for the mean force, i.e., $m=0$, read 
\begin{align}
    \Gamma_{s}^{(0,1)}  &= \beta^2 \langle F_{s};F_0\rangle_\mathrm{eq}, \label{eq:Gamma1general} \\
     \Gamma_{s,s'}^{(0,2)} &=  \frac{\beta^2}{2} \left( \left\langle \mathcal{D}_{s};F_{s'};F_0 \right\rangle_\mathrm{eq} + \langle \mathcal{D}_{s'};F_{s};F_0\rangle_\mathrm{eq} \right),\label{eq:Gamma2general} 
     \\
      \begin{split}
    \Gamma_{s_1,s_2,s_3}^{(0,3)} &= \frac{1}{6} \sum_{\pi \in S_3} \Bigg(\frac{\beta^2}{2} \Big[ \left\langle \mathcal{D}_{s_{\pi(1)}};\mathcal{D}_{s_{\pi(2)}}; F_{s_{\pi(3)}}; F_0 \right\rangle_\mathrm{eq} \\
    &\quad \quad\quad \quad \quad \quad - \left\langle \mathcal{D}_{s_{\pi(1)},s_{\pi(2)}};F_{s_{\pi(3)}};F_0 \right\rangle_\mathrm{eq} \Big]\\
    &\quad \quad\quad \quad + \frac{\beta^4}{24} \langle F_{s_{\pi(1)}}; F_{s_{\pi(2)}}; F_{s_{\pi(3)}}; F_0\rangle_\mathrm{eq}\Bigg),\label{eq:Gamma3general}
    \end{split}\\
        &\;\;\vdots  \nonumber
\end{align}
where the sum in $\Gamma^{(0,3)}$ goes over all elements of the permutation group $S_3$.
The first kernels entering the force covariance $\langle F_t;F_{t'}\rangle$, i.e., $m=1$ are given by 
\begin{align}
       \Gamma^{(1,0)}_{t-t'} &= \beta^2 \langle F_{t-t'};F_0\rangle_\mathrm{eq}, \label{eq:Gamma10general}\\
        \Gamma^{(1,1)}_{s;t-t'} &= \beta^2 \left\langle \left( \mathcal{D}_{s} + \frac{\beta F_{s}}{2} \right);F_{t-t'};F_0\right\rangle_\mathrm{eq}, \label{eq:Gamma11general} 
\end{align}
    \begin{align}
         \begin{split}
        \Gamma^{(1,2)}_{s_1,s_2;t-t'} &= \frac{\beta^2}{4} \sum_{\pi \in S_2} \Bigg[ \langle \mathcal{D}_{s_{\pi(1)}};\mathcal{D}_{s_{\pi(2)}};F_{t-t'};F_0\rangle_\mathrm{eq} \\
        &\quad\quad\quad\quad- \langle \mathcal{D}_{s_{\pi(1)},s_{\pi(2)}};F_{t-t'};F_0\rangle_\mathrm{eq} \\
        &\quad\quad\quad\quad + \beta \langle \mathcal{D}_{s_{\pi(1)}};F_{s_{\pi(2)}};F_{t-t'};F_0\rangle_\mathrm{eq} \\
        &\quad\quad\quad\quad+ \frac{\beta^2}{4} \langle F_{s_{\pi(1)}};F_{s_{\pi(2)}};F_{t-t'};F_0\rangle_\mathrm{eq} \Bigg].
        \end{split}\label{eq:Gamma12general}\\
        &\;\;\vdots  \nonumber
        \end{align}
        These thus quantify how the force covariance changes due to the driving. The kernels for the third cumulant $\langle F_t;F_{t'};F_{t''}\rangle$ ($m=2$) read
        \begin{align}
        \Gamma^{(2,0)}_{t-t',t-t''} &= \beta^3 \langle F_{t-t'};F_{t-t''};F_0\rangle_\mathrm{eq}, \label{eq:Gamma20general} \\
        \Gamma^{(2,1)}_{s;t-t',t-t''} &= \beta^3\left\langle \left( \mathcal{D}_{s}+\frac{\beta F_{s}}{2}  \right) ;F_{t-t'};F_{t-t''};F_0\right\rangle_\mathrm{eq}. \label{eq:Gamma21general}\\
        &\;\;\vdots  \nonumber
        \end{align}
The kernels with $m=2$ are thus the leading measure how non-Gaussianity occurs due to, or is affected by driving.

Due to the determinacy of the protocol $\dot{x}_t$, $x_t$ can be moved out of averages, and the kernels only contain forces $F_s$ and NT coefficients $\mathcal{D}_s$, $\mathcal{D}_{s,s'}$.
We have applied time-reversal symmetry and stationarity of equilibrium correlation functions, thereby finding that all kernels $\Gamma^{(m,n)}_{t-s_1,\dots,t-s_n;t-t_1,\dots,t-t_m}$ only depend on time differences $t-s_1$, $t-s_2$, \dots, as already assumed in  Eqs.~\eqref{eq:MeanForce} -  \eqref{eq:ForceCum3}.
From the definition, Eq.~\eqref{eq:Gammamn}, it becomes clear that the  kernels are separately symmetric under exchange of the time arguments within one block, i.e.~under changes of time arguments $s_1,\dots,s_n$ and, separately, under changes of time arguments $t_1,\dots,t_m$. 
The symmetry with respect to changes of time arguments $s_i$ is enforced by hand, namely through the sums over all elements of respective permutation groups in Eqs.~\eqref{eq:Gamma1general}-\eqref{eq:Gamma21general}. 
We note that Eqs.~\eqref{eq:Gamma10general} and Eqs.~\eqref{eq:Gamma1general} display the equality  $\Gamma^{(0,1)}= \Gamma^{(1,0)}$, which is the fluctuation dissipation theorem.

We derive a general expression for the $n-$th order memory kernel $\Gamma^{(0,n)}$ in Appendix~\ref{app:Gamman}. This expression uses Eq.~\eqref{eq:ResponseStateO}, i.e., it is not given in terms of connected correlation functions.
We thus provide a framework to study non-Gaussian fluctuations in non-equilibrium.

\section{Interpretation of nonlinear Langevin functionals}\label{chap:Embedding}

We have in the introduction mentioned that nonlinear stochastic motion is conceptually and practically challenging, with intense efforts and developments \cite{glatzel_interplay_2021,vroylandt_position-dependent_2022,meyer_non-stationary_2017,te_vrugt_mori-zwanzig_2019,meyer_dynamics_2019,netz_derivation_2023}.    

How are Eqs.~\eqref{eq:MeanForce}-\eqref{eq:Gammamn} interpreted? These are the non-equilibrium force cumulants, which  characterize the non-Gaussian, non-equilibrium statistics of the force acting on an externally driven control parameter $x_t$. This is relevant and experimentally observable, for example via a Brownian particle confined in a moving trapping potential \cite{wilson_small-world_2011,berner_oscillating_2018,jain_two_2021}.  

It is however important to keep in mind that $\dot x$ appearing in Eqs.~\eqref{eq:MeanForce}-\eqref{eq:Gammamn} is the deterministic velocity of the control parameter. Our formulation is thus nonlinear in this control parameter. It is not a nonlinear equation for a stochastic degree of freedom. This formulation is thus, a priori, not an equation of motion of a Brownian particle. It may however, in certain scenarios or limits, for example the Markovian limit, turn into a nonlinear equation for a stochastic degree. This is not known to us and needs to be investigated in future work.   

What we derive here is thus not a Langevin {\it equation}, and we refer to Eqs.~\eqref{eq:MeanForce}-\eqref{eq:ForceCum3} as nonlinear Langevin {\it functionals}.

\section{Driving in the stochastic Prandtl-Tomlinson model}\label{chap:SteadyShearSPT}

\subsection{Introduction and protocol}

In this section we will discuss simulation results for a specific system of Brownian overdamped degrees of freedom, the so-called stochastic Prandtl-Tomlinson model~\cite{muller_properties_2020,jain_two_2021,jain_micro-rheology_2021}.
In previous works, this model was shown to contain several nonlinear phenomena experimentally observed in viscoelastic systems such as shear-thinning, particle oscillations~\cite{jain_two_2021} and a trap stiffness dependent memory kernel~\cite{muller_properties_2020}.
We will study the nonlinear force functionals for two driving modes: \textit{direct} and \textit{indirect driving}. These correspond to two different controlled degrees of freedom as will be detailed below.

We restrict to steady driving velocity, i.e., a protocol of
\begin{align}
    \dot x_t=v \Theta(t-t_0),\label{eq:proto}
\end{align}
and we denote the corresponding averages by $\langle\dots\rangle^{(-t_0)}$. Using $t_0=0$ and  $t_0\to-\infty$ yields transient and steady state averages, respectively. 
\subsection{Model and driving modes}

\begin{figure}
    \centering
    \includegraphics{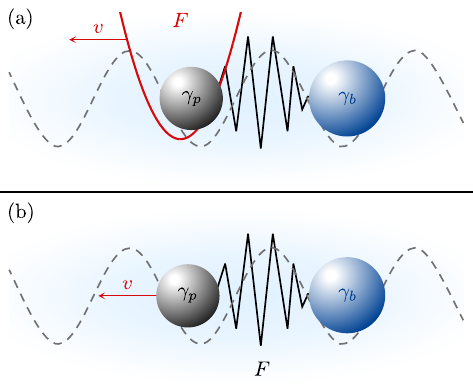}
    \caption{Sketch of the Stochastic Prandtl-Tomlinson model: a probe particle (gray) with friction coefficient $\gamma_p$ is coupled to a bath particle (blue) with friction coefficient $\gamma_b$ via a sinusoidal interaction potential (dashed line). The upper panel (a) highlights the scenario of indirect driving, in which the probe is confined in a harmonic trapping potential (red), i.e.~experiences a trapping force $F$. The center of the trapping potential is driven with velocity $v$. The lower panel (b) shows the scenario of direct driving in which the position of the probe particle corresponds to the protocol, hence follows a given velocity $v$. In this scenario, the force $F$ of the laid out formalism is the interaction force between probe and bath. }
    \label{fig:SPTSketch}
\end{figure}

Consider two coupled Brownian particles: one probe particle at position $y^{(p)}$ and one bath particle at position $y^{(b)}$.
Each of the particles is characterized by a friction coefficient $\gamma_p$ or $\gamma_b$, respectively.
The two particles interact via a sinusoidal potential~\cite{muller_properties_2020} 
\begin{align}
    U_\mathrm{int}\left(y^{(b)}-y^{(p)}\right) = - U_0 \cos \left( \frac{2 \pi}{d} \left(y^{(b)}-y^{(p)}\right) \right)
    \label{eq:SPTPotential}
\end{align}
with amplitude $U_0$ and length scale $d$, see dashed line in Fig.~\ref{fig:SPTSketch}.
The interaction potential is thus a non-binding potential, such that probe and bath particle can move with different (mean) velocities, when the probe is driven~\cite{muller_properties_2020}.  
This is for example required to observe shear-thinning~\cite{jain_micro-rheology_2021,jain_two_2021}. This model was termed stochastic Prandtl-Tomlinson model \cite{muller_properties_2020}.

In \textit{indirect driving}, the probe particle is confined in a trapping potential, whose center is the controlled parameter $x_t$, see Fig.~\ref{fig:SPTSketch}(a). 
In this mode, $y^{(1)} = y^{(p)}$ and $y^{(2)} = y^{(b)}$ are the two Brownian overdamped degrees of freedom with $\gamma_1 = \gamma_p$ and $\gamma_2 = \gamma_b$ that obey Eq.~\eqref{eq:MasterLangevin}. 
Here, we choose a harmonic trapping potential $U_\mathrm{ext}\left(y^{(1)}-x\right) = \frac{1}{2} \kappa \left(y^{(1)}-x\right)^2$.
In \textit{direct driving}, the position of the probe particle is deterministic and follows the protocol, i.e.~$y^{(p)} =x$ (Fig.~\ref{fig:SPTSketch}(b)). 
In this driving mode we have one Brownian overdamped degree of freedom, the position of the bath particle, i.e.~$y^{(1)} = y^{(b)}$ and $\gamma_1 = \gamma_b$.

The two driving modes correspond to the following potential  $U = U(\{y^{(i)}-x\})$ in Eq.~\eqref{eq:MasterLangevin}, 
\begin{align}
    U
    = \begin{cases}
        U_\mathrm{int}\left(y^{(2)}-y^{(1)}\right) + U_\mathrm{ext}\left(y^{(1)}-x\right), & \text{indirect} \\
        U_\mathrm{int}\left(y^{(1)}-x\right), &\text{direct}.
    \end{cases}
\end{align}
The resulting force $F_t = \partial_{x_t} U(\{y^{(i)}_t-x_t\})$ reads in the two cases
\begin{align}
    F_t = \begin{cases}
        - \kappa q_t^{(1)}, & \text{indirect driving} \\
        - \frac{2 \pi}{d} U_0 \sin \left( \frac{2 \pi}{d} q_t^{(1)} \right), & \text{direct driving},
    \end{cases}\label{eq:FtSPT}
\end{align}
expressed in terms of the relative coordinates $q^{(i)} = y^{(i)}-x$.
We follow the protocol in Eq.~\eqref{eq:proto}, and we fix 
 the choices $\gamma_b = 10 \,\gamma_p$ and $U_0 = 2\, k_B T$ throughout: Using  $\gamma_b$ large compared to $\gamma_p$ yields pronounced nonlinear effects. We have found that $U_0$ slightly larger than thermal energy is the most interesting regime of the Prandtl Tomlinson model \cite{jain_micro-rheology_2021}: For   $U_0 \ll  k_B T$, the effects of the coupling vanish, and for $U_0 \gg  k_B T$, the coupling becomes nearly that of a harmonic spring. 
\subsection{1st Cumulant: Flow Curve}\label{chap:SPT_Flowcurve}
We start by considering the mean force, i.e., we evaluate Eq.~\eqref{eq:MeanForce} in steady state conditions.
Dividing the mean force by the driving velocity yields the velocity dependent friction coefficient 
$\gamma(v)$, or flow curve~\cite{jain_two_2021}.
For the two scenarios of direct and indirect driving we define~\cite{jain_two_2021,muller_brownian_2020}
\begin{align}
    \gamma(v) = \frac{\left|\langle F\rangle^{(\infty)}\right|}{v} + \begin{cases}
        0, & \text{indirect driving} \\
        \gamma_p, & \text{direct driving}.
    \end{cases}\label{eq:Gamma(v)}
\end{align}
For a better comparison between the two scenarios we added $\gamma_p$ to the case of direct driving in Eq.~\eqref{eq:Gamma(v)}. With this, $\gamma(v)$ of the two cases approach each other for $\kappa\to\infty$, as we shall see below. 

From Eq.~\eqref{eq:MeanForce} we find for $\langle F\rangle^{(\infty)}$, 
\begin{align}
    \beta\frac{\left|\langle F\rangle^{(\infty)}\right|}{v}  = \hat{\Gamma}_{z=0}^{(0,1)} + 
   v^2 \hat{\Gamma}_{z=0,z=0,z=0}^{(0,3)} + \mathcal{O}(v^4)\label{eq:FlowcurveFromLangevin}
\end{align}
with Laplace transforms defined as $\hat{f}_z = \int_0^\infty \mathrm{d}t \, e^{-z t} f_t$; with definitions for multiple time dependencies accordingly.
We note that  $\Gamma^{(0,2)}$ misses in Eq.~\eqref{eq:FlowcurveFromLangevin} as it is zero due to symmetries. The flow curve is  an expansion in even powers of $v$.

In Appendix~\ref{app:ConvergenceConnectedCorr} we discuss in detail how to efficiently compute time-integrated memory kernels, considering the example of $\hat{\Gamma}^{(0,3)}_{z=0,z=0,z=0}$. 
Moreover, we compare two different ways to do this, based on (i) the "naive" response relation, Eq.~\eqref{eq:ResponseStateO}, and (ii) the response relation in terms of connected correlation functions, Eq.~\eqref{eq:ResponseStateOConnectedCorrelations}.
As mentioned above, connected correlation functions are advantageous as they analytically cancel divergences in individual response terms. 

\begin{figure}
    \centering
    \includegraphics{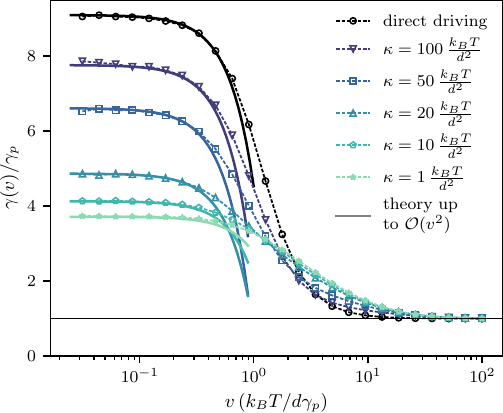}
    \caption{Flow curve $\gamma(v)$ (open symbols with dashed lines) and response curves up to second order according to Eq.~\eqref{eq:FlowcurveFromLangevin} (solid lines) for direct driving (black) and indirect driving (colors) with velocity $v$. Different colors correspond to different stiffness $\kappa$ of the harmonic potential. 
    }
    \label{fig:flowcvComparison}
\end{figure}

Fig.~\ref{fig:flowcvComparison} shows the resulting comparison of Eq.~\eqref{eq:Gamma(v)} computed from non-equilibrium simulations (symbols with dashed lines) and  response curves up to third order in driving velocity $v$ using equilibrium simulations (according to Eq.~\eqref{eq:FlowcurveFromLangevin}).
The figure shows flow curves for different stiffness $\kappa$ of the trapping potential (colors) and for direct driving (black).
All flow curves exhibit  shear-thinning,
as extensively discussed for the stochastic Prandtl-Tomlinson model in Refs.~\onlinecite{jain_micro-rheology_2021,muller_brownian_2020}.
We find that shear-thinning, as might be expected, indeed  appears in the response curves (solid lines) up to third order. 
The first-order kernel $\hat{\Gamma}^{(0,1)}_{z=0}$ (with $\gamma_p$ added in the direct case) characterizes the plateau $\gamma(v\to 0)$ and increases monotonously with $\kappa$; the probe experiences stronger interaction with the bath particle, if the confinement is strong~\cite{muller_properties_2020}.
The third-order kernel $\hat{\Gamma}_{z=0,z=0,z=0}^{(0,3)}$ characterizes the onset of shear-thinning and shows a non-monotonous behavior in $\kappa$. This is shown in Fig.~\ref{fig:Gamma_vs_kappa} (discussed in Sec.~\ref{chap:commutation}), displaying that, when the trap stiffness matches the curvature of the interaction potential, nonlinear effects set in most early as a function of $v$.

\subsection{2nd Cumulant: Force fluctuations}\label{chap:ForceFluctuations}

\begin{figure}
    \includegraphics{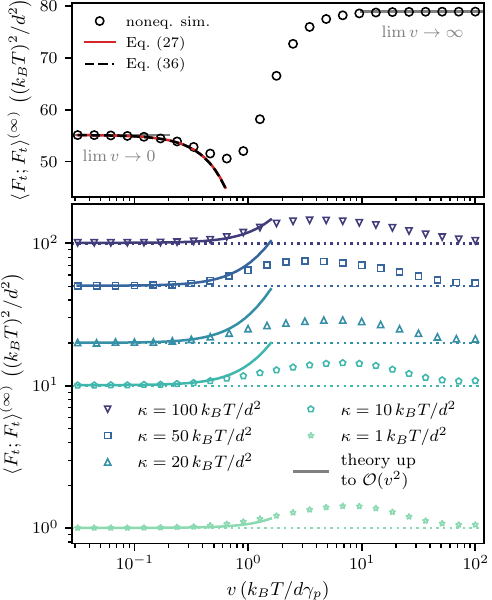}
    \caption{Steady-state force variance as a function of driving velocity for the scenario of direct (top) and indirect (bottom) driving. Open symbols show results from non-equilibrium simulations, the lines correspond to response curves up to $\mathcal{O}(v^2)$. In the top panel two response curves according to Eq.~\eqref{eq:ForceCov} using Eq.~\eqref{eq:ForceCovEqualTimes} (black dashed line) and Eq.~\eqref{eq:Gamma12general} (red line) are shown, demonstrating that they yield identical results as expected.
    The gray lines depict the limits $v \to 0$ and $v\to \infty$ according to Eqs.~\eqref{eq:ForceVarianceEqDirect} and \eqref{eq:ForceVarianceVInfDirect}.
    For the scenario of indirect driving the limit of $v \to 0$ given by Eq.~\eqref{eq:ForceVarianceEqIndirect} is indicated by the colored dotted lines in the bottom panel.}
    \label{fig:ForceCovEqualTimes}
\end{figure}

We continue by exploiting Eq.~\eqref{eq:ForceCov} to study the force covariance $\langle F_t;F_{t'}\rangle^{(-t_0)}$ in various cases, starting with steady-state, i.e., $t_0\to -\infty$ and $t=t'$. For this case, the second order memory kernel in Eq.~\eqref{eq:ForceCov} can be simplified to
\begin{align}
\begin{split}
   \Gamma^{(1,2)}_{s_1,s_2;0} = \frac{\beta^2}{2} \sum_{\pi \in S_2} \left\langle \mathcal{D}_{s_{\pi(1)}};F_{s_{\pi(2)}};F_0;F_0\right\rangle_\mathrm{eq} ,
    \end{split}\label{eq:ForceCovEqualTimes}
\end{align}
compare Eq.~\eqref{eq:ResponseStateOConnectedCorrelations}.
In Fig.~\ref{fig:ForceCovEqualTimes} we explore the force variance as a function of driving velocity for the two scenarios of direct (top panel) and indirect (lower panel) driving.
Open symbols show results from non-equilibrium simulations; lines show response curves up to second order in driving velocity obtained from equilibrium simulations.

In the scenario of direct driving (Fig.~\ref{fig:ForceCovEqualTimes}, top panel) the force variance converges to plateaus in the two limiting cases of $v \to 0$ and $v \to \infty$ and exhibits a minimum in between. 
The limits $v \to 0$ and $v \to \infty$ can be found analytically, see gray lines in the top panel in Fig.~\ref{fig:ForceCovEqualTimes}.
In equilibrium the relative coordinate $q^{(1)}$ between bath and probe particle obeys a Boltzmann distribution with the interaction potential $U_\mathrm{int}(q^{(1)})$ (Eq.~\eqref{eq:SPTPotential}) which yields
\begin{align}
  \lim_{v \to 0} \langle F_t;F_t\rangle^{(\infty)}=  \langle F_t;F_t\rangle_\mathrm{eq} = \frac{4 \pi^2 U_0 I_1(\beta U_0)}{\beta d^2 I_0(\beta U_0)},\label{eq:ForceVarianceEqDirect}
\end{align}
where $I_n(z)$ denotes the Bessel function of first kind.
In the limit of infinite velocity the distribution of the distance between probe and bath particle can be assumed flat, such that
\begin{align}
\begin{split}
    \lim_{v \to \infty} \langle F_t;F_t\rangle^{(\infty)} &= \int_0^d \mathrm{d}q^{(1)}\left(\frac{2 \pi U_0}{d}\sin\left( \frac{2 \pi }{d} q^{(1)}\right)\right)^2\\
    &= \frac{2 \pi^2 U_0^2}{d^2},
    \end{split}
    \label{eq:ForceVarianceVInfDirect}
\end{align}
which is also shown in the figure. Notably, $\lim_{v \to \infty} \langle F_t;F_t\rangle^{(\infty)} \geq \langle F_t;F_t\rangle_\mathrm{eq}$ for any $\beta U_0$, with equality being approached for $\beta U_0\to0$. 

The behavior for finite velocity can be understood in the following picture: with increasing driving velocity the probe experiences the anharmonic part of the interaction potential ($\sim -x^4$) and the effective curvature decreases, resulting in a decay of the force variance.
For larger driving velocities, when shear-thinning takes place, the probe can hop over the barrier and explore the full range of forces, so that the force variance increases.
We find that the nonlinear phenomenon of initial decay of the force variance appears in the second-order kernel of the force covariance, $\Gamma^{(1,2)}$. 
Moreover, we confirm Eq.~\eqref{eq:ForceCovEqualTimes} by comparing response curves using Eq.~\eqref{eq:ForceCovEqualTimes} (dashed black line) and using the original expression, Eq.~\eqref{eq:Gamma12general} (red line) in Fig.~\ref{fig:ForceCovEqualTimes} (top panel). 

In the scenario of indirect driving (Fig.~\ref{fig:ForceCovEqualTimes}, bottom panel) we observe a maximum at finite driving velocity for all different values of trap stiffness $\kappa$ (colors).
This is feature has been characterized by an increased effective temperature~\cite{cugliandolo_effective_2011,puglisi_temperature_2017,wilson_small-world_2011}.
Also here, the initial increase is captured by the second order memory kernel $\Gamma^{(1,2)}$.
The dotted lines in the bottom panel in Fig.~\ref{fig:ForceCovEqualTimes} correspond to the limit $v \to 0$, found from the Boltzmann distribution with the harmonic trapping potential,
\begin{align}
   \lim_{v \to 0} \langle F_t;F_t\rangle^{(\infty)} =\langle F_t;F_t\rangle_\mathrm{eq} = \kappa k_B T. \label{eq:ForceVarianceEqIndirect}
\end{align}
We observe that in the limit of infinite velocity, when probe and bath particle are decoupled, the data points of $\langle F_t;F_t\rangle^{(\infty)}$ appear to approach the dotted line of equilibrium values. This may be understood as the interaction with the bath particle acts as a fixed periodic potential background in this limit. When the probe particle is dragged over this background, the force thus rapidly switches sign and cancels out.
Notably, the force fluctuations within the two scenarios 
are inherently different: 
in the scenario of direct driving they characterize the interaction between probe and bath out of equilibrium, while they characterize the statistics of the trapping force in the scenario of indirect driving.\\

\begin{figure}
    \includegraphics{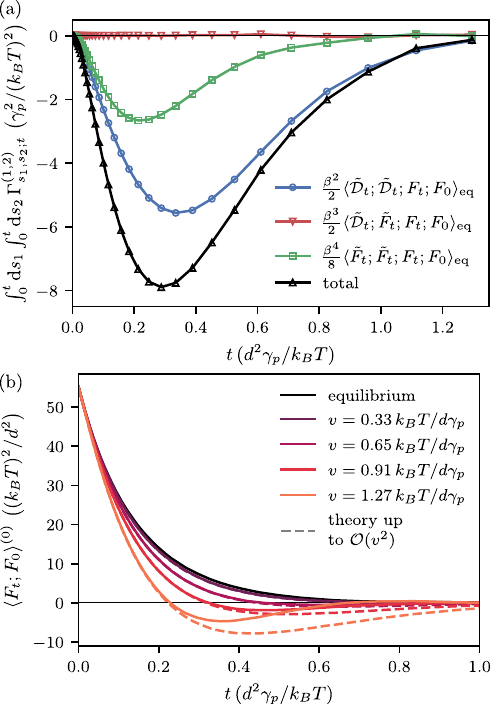}
    \caption{(a) Time-dependence of second order response terms for the transient force covariance in the scenario of direct driving according to Eq.~\eqref{eq:ForceCov}. Note that $\tilde{f}_t = \int_0^t \mathrm{d}s\, f_s$ is defined as the running integral up to time $t$. (b) Resulting comparison of the transient force covariance (solid lines) with the theoretical response curves up to $\mathcal{O}(v^2)$ (dashed lines) according to Eq.~\eqref{eq:ForceCov} for several driving velocities in the scenario of direct driving.}
    \label{fig:TransientForceCovariance}
\end{figure}

Next, we study Eq.~\eqref{eq:ForceCov} for $t'=0$ and finite $t$, i.e.~consider the transient force covariance $\langle F_t;F_0\rangle^{(0)}$ where the system is prepared in equilibrium at time $t=0$.
Therefore, we require the time-dependence of the second-order memory kernel, Eq.~\eqref{eq:Gamma12general}.
The resulting curves for the scenario of direct driving
are shown in Fig.~\ref{fig:TransientForceCovariance}(a) and exhibit a minimum at short times.
Adding the black curve times velocity squared to the equilibrium force covariance $\langle F_t;F_0\rangle_\mathrm{eq}$ we obtain the response curves according to Eq.~\eqref{eq:ForceCov}, that we can compare to the transient force covariance for several driving velocities.
This comparison is shown in Fig.~\ref{fig:TransientForceCovariance}(b), where solid lines correspond to the transient force covariance obtained from non-equilibrium simulations and the dashed lines give response curves calculated from equilibrium simulations.
While in equilibrium the force covariance monotonously decays to zero, we observe a minimum for finite driving velocities. 
Furthermore, $\langle F_t;F_0\rangle^{(0)} $ is negative for intermediate times.
This phenomenon has been discussed extensively in previous  works for indirect driving~\cite{berner_oscillating_2018,jain_two_2021,jain_micro-rheology_2021}. Here, we also observe it for direct driving, which awaits experimental testing. The case of indirect driving shows the same behavior, i.e., the previously found oscillations \cite{berner_oscillating_2018} emerge in third order, see Fig.~\ref{fig:TransientForceCovKappa100} in Appendix~\ref{app:OscillationsIndirectDriving}. 

We observe that, in this case, the convergence of the Volterra series also depends on time $t$. The response curve, for the transient case, is by construction   exact for  $t = 0$ for all driving velocities (as this corresponds to the equilibrium case), and the deviation naturally increases with increasing $t$ for a given fixed $v$, as seen in Fig.~\ref{fig:TransientForceCovariance}.

\subsection{3rd Cumulant: Non-Gaussian force fluctuations}\label{chap:Non-GaussianFluctuations}

\begin{figure}
    \centering
    \includegraphics{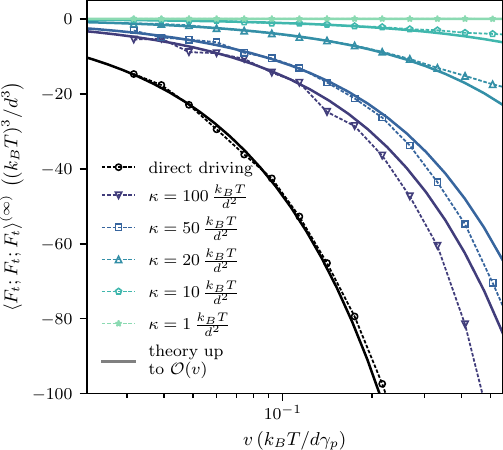}
    \caption{Third force cumulant as a function of driving velocity for the scenario of direct (black) and indirect driving (colors). Open symbols with dashed lines show results from non-equilibrium simulations. Lines correspond to  response curves up to linear order obtained from equilibrium simulations (according to Eq.~\eqref{eq:ForceCum3}). In the limit $v \to 0$ all curves converge to zero due to the symmetries of the equilibrium case.}
    \label{fig:ForceCum3}
\end{figure}

Higher-order force cumulants quantify non-Gaussian fluctuations.
As an example we consider the steady-state third force cumulant $\langle F_t;F_t;F_t\rangle^{(\infty)}$ evaluated at equal times, found from Eq.~\eqref{eq:ForceCum3}.
Due to the equal times the expression for the respective first-order memory kernels simplifies to 
\begin{align}
    \Gamma^{(2,1)}_{s;0,0} = \beta^4 \left\langle F_s;F_0;F_0;F_0 \right\rangle_\mathrm{eq}.
\end{align}
Fig.~\ref{fig:ForceCum3} shows $\langle F_t;F_t;F_t\rangle^{(\infty)}$ as a function of driving velocity for  direct driving (black) and indirect driving for different stiffness $\kappa$ (colors). 
We compare curves from non-equilibrium simulations (open symbols with dashed lines) with theoretical response curves up to linear order according to Eq.~\eqref{eq:ForceCum3} (solid lines).
In the limit $v \to 0$ the third force cumulant vanishes for both scenarios due to symmetry; the force distributions are symmetric in equilibrium.
For finite driving Fig.~\ref{fig:ForceCum3} shows non-Gaussian behavior, whose amplitude increases with $\kappa$ and is particularly strong in the scenario of direct driving.
We attribute this to the fact that in the scenario of direct driving  the interaction potential  is nonlinear.

\subsection{Non-commutation of limits $\kappa \to \infty$ and $t \to 0$}\label{chap:commutation}

\begin{figure}
    \centering
    \includegraphics{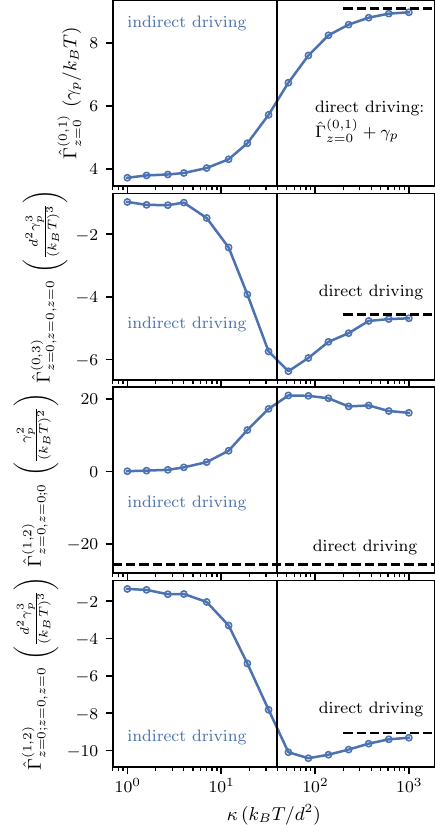}
    \caption{Integrated memory kernels $\hat{\Gamma}_{z=0}^{(0,1)}$ (top), $\hat{\Gamma}^{(0,3)}_{z=0,z=0,z=0}$ (second), $\hat{\Gamma}^{(1,2)}_{z=0,z=0;0}$ (third) and $\hat{\Gamma}^{(1,2)}_{z=0;z=0,z=0}$ (bottom) of mean force $\beta \langle F_t\rangle^{(\infty)}$, force variance $\beta^2 \langle F_t;F_t\rangle^{(\infty)}$ and integrated force covariance $\beta^2 \langle \hat{F}_{z=0};F_0\rangle^{(\infty)}$ as a function of $\kappa$ for the scenario of indirect driving. The black dashed lines show resulting values for the scenario of direct driving. 
    In the first panel $\gamma_p$ has been added to $\hat{\Gamma}^{(0,1)}_{z=0}$ for direct driving.
    The vertical line corresponds to $\kappa = \frac{2 \pi^2 U_0}{d^2} \approx 39 k_B T/d^2$, half the curvature of the interaction potential $U_\mathrm{int}$ in the minimum. }
    \label{fig:Gamma_vs_kappa}
\end{figure}

We have in Secs.~\ref{chap:SPT_Flowcurve}-\ref{chap:Non-GaussianFluctuations} observed that in some cases, the limit of $\kappa\to\infty$ agrees with the case of direct driving, and in some cases, it does not. In this subsection, we analyze this in more detail, along with Fig.~\ref{fig:Gamma_vs_kappa}.

In the limit of $\kappa\to\infty$, the probe particle follows a prescribed velocity, so that the scenarios are expected to be identical. Furthermore, the force balance between bath particle and probe particle implies that the first cumulant, i.e., the mean force, agrees in that limit in the two cases. Indeed, the flow curve $\gamma(v)$ for indirect driving approaches the one of direct driving in the limit of $\kappa\to\infty$, see the two upper panels in  Fig.~\ref{fig:Gamma_vs_kappa};  $\hat{\Gamma}_{z=0}^{(0,1)}$ and $\hat{\Gamma}^{(0,3)} _{z=0,z=0,z=0}$ are the leading contributions to the flow curve for small $v$ (compare Eq.~\eqref{eq:FlowcurveFromLangevin}, with $\gamma_p$ added in the direct case), and the graphs show that the cases of indirect driving reaches the direct case for large $\kappa$.

The situation is different for the force variance. Here, even in equilibrium, $\langle F_t; F_t\rangle_{\rm eq}$ in Eq.~\eqref{eq:ForceVarianceEqDirect} does not approach the case of direct driving of Eq.~\eqref{eq:ForceVarianceEqIndirect} for large $\kappa$ (as observed in Fig.~\ref{fig:ForceCovEqualTimes}). 
We recall that this difference in the force variance is an interesting feature showing that the direct case may  allow experimental investigation of the interaction potential between particle and its surrounding. Also for finite velocity, $\langle F_t ;F_t\rangle^{(\infty)}$ for the indirect case does not approach the direct case for large $\kappa$, as demonstrated in Fig.~\ref{fig:ForceCovEqualTimes} and in the third panel of Fig.~\ref{fig:Gamma_vs_kappa}; $\hat{\Gamma}^{(1,2)}_{z=0,z=0;0}$ describes the leading non-equilibrium contribution to that quantity, and, as seen, the two cases do not agree for large $\kappa$. 

Does this observation of the limit of $\kappa\to\infty$ not reaching the case of direct driving violate the mentioned force balance? This might be understood by using the equilibrium case,  Eqs.~\eqref{eq:ForceVarianceEqDirect} and  \eqref{eq:ForceVarianceEqIndirect}. These two equations result of equilibrium Boltzmann distributions of two distinct potentials, the harmonic trapping potential on one side, and the nonlinear interaction potential on the other. These thus lead to distinctly different results for the variance. The force balance comes however into play for finite time $t$ in $\langle F_t ;F_0\rangle^{(\infty)}$. If $\kappa$ is large, we observe the force covariance for the indirect case to follow $\langle F_t ;F_0\rangle^{(\infty)}\propto e^{-\frac{\kappa}{\gamma_p}t}$ for $t \ll \gamma_p/\kappa$, and to agree with the result of the direct case for $t \agt \gamma_p/\kappa$. In other words the time period, where the two functions disagree, gets shorter and shorter, as $\kappa$ increases. We can thus state that limits of $\kappa\to\infty$ and $t\to 0$ seem not to commute. Setting $t= 0$ in  $\langle F_t ;F_0\rangle^{(\infty)}$ yields no agreement of the two cases for any $\kappa$, while, for any finite $t$, the two cases approach each other in the limit of $\kappa\to\infty$.

This can be illustrated by regarding the time integral, $\int_0^\infty \mathrm{d}t\, \langle F_t ;F_0\rangle^{(\infty)}$. For $v=0$, i.e., in equilibrium, this equals the $v\to 0$ limit of the flow curve, via FDT; thus, indeed, the cases approach each other in the limit of  $\kappa\to\infty$ (taking into account the addition of $\gamma_p$ in the direct case). The leading non-equilibrium contribution to $\int_0^\infty \mathrm{d}t\, \langle F_t ;F_0\rangle^{(\infty)}$ is given by the integrated kernel $\hat{\Gamma}^{(1,2)}_{z=0;z=0,z=0}$, shown in the lowest panel of Fig.~\ref{fig:Gamma_vs_kappa}. We see that the two cases indeed approach each other for $\kappa\to\infty$. 

The non-commutativity of the limits $\kappa \to \infty$ and $t \to 0$ can be further illustrated by comparing Figs.~\ref{fig:TransientForceCovariance}(b) and \ref{fig:TransientForceCovKappa100} in Appendix~\ref{app:OscillationsIndirectDriving}, i.e.~the transient force covariance in the two scenarios, using a quite large value of $\kappa=100 \,k_BT/d^2$.
The two graphs are similar despite an additional fast initial decay in the scenario of indirect driving, following the mentioned behavior $\langle F_t;F_0\rangle \approx \kappa k_B Te^{-\frac{\kappa}{\gamma_p}t}$.
For longer times the curves for the two scenarios fall on top of each other. 

We conclude with another interesting aspect of Fig.~\ref{fig:Gamma_vs_kappa}.
The graphs include vertical lines marking the point where $\kappa$ equals half the curvature of the interaction potential, i.e.,  $\kappa = \frac{1}{2} U_{\rm int}''(0)= \frac{2 \pi^2 U_0}{d^2}$. Interestingly, the curves show pronounced changes around this value. We may thus expect that, in experiments, the kernels strongly depend on $\kappa$ when $\kappa$ is comparable to the curvature of the interaction potential with the surrounding.

\section{Conclusion}

We derive nonlinear Langevin functionals for the cumulants of the stochastic force acting on a driven controlled degree of freedom, in a surrounding of Brownian particles. These nonlinear Langevin funtionals, Eqs.~\eqref{eq:MeanForce}-\eqref{eq:ForceCum3}, are of the form of Volterra series in the velocity of the control parameter. We provide them explicitly up to third order in driving velocity, and highlight 
that respective kernels are given in terms of {\it connected} correlation functions; These follow from general nonlinear response formulas in terms of connected correlation functions derived here, and which we find to cancel divergences in the examples investigated. 

We highlighted and discussed the specific scenarios of direct and indirect driving of a Brownian probe, using simulations of the stochastic Prandtl-Tomlinson model to explicitly evaluate and verify the kernels of Eqs.~\eqref{eq:MeanForce}-\eqref{eq:ForceCum3} in several limiting cases for the two scenarios. We find that qualitative nonlinear phenomena, such as shear-thinning and oscillations in the force covariance already appear in the memory kernels of third or second order.

We find that the force variance reveals interesting differences between the two cases of direct and indirect driving, and in particular, the direct case allows to infer information about the interaction potential of probe and surrounding. When regarded as a function of time difference $t$, this variance appears to be continuous in $\kappa$ for finite time $t$, meaning that the indirect case approaches the direct one for $\kappa\to\infty$, while the two cases remain different at $t=0$. 
We interpret that the limits of $\kappa \to \infty$ and $t \to 0$ do not commute, an effect which becomes apparent in the force fluctuations, but not in the mean force.

With Eqs.~\eqref{eq:MeanForce}-\eqref{eq:Gammamn} we have established a complete theoretical framework for studying nonlinear and non-equilibrium fluctuations in overdamped Brownian baths, which opens up rich possibilities for further investigations, for example different driving protocols or coupling of different driven degrees of freedom, i.e., multidimensional controlled degrees. It will also be important to find whether the series expansion in terms of connected correlation functions can be extended to higher orders. One may also investigate whether time derivatives appearing in the correlation functions can be replaced~\cite{asheichyk_response_2019}.

\section*{Acknowledgments}
This project was funded by the Deutsche Forschungsgemeinschaft (DFG), Grant No. SFB 1432 (Project ID 425217212) - Project C05. We thank Clemens Bechinger, Luis Reinalter, Michael Jade-Jerez and Kiryl Asheichyk for discussions. 

\section*{Author declarations}
\subsection*{Conflict of interest}
The authors have no conflicts to disclose.

\subsection*{Author Contributions}
\noindent \textbf{Juliana Caspers}: Formal Analysis, Investigation, Software, Visualization, Writing - Original Draft. \\
\noindent \textbf{Matthias Krüger}: Conceptualization, Methodology, Writing – Review \& Editing.

\section*{Data availability}
The data that support the findings of this study are available from the corresponding author upon reasonable request.

\appendix

\section{Connected correlation functions}\label{app:connectedCorr}

The response relations Eqs.~\eqref{eq:ResponseO} and \eqref{eq:ResponseStateO} must hold for any state observable $O_t$, e.g.~also for $O =1$.
We can thus retrieve a list of identities that entropic and NT components must obey,
\begin{align}
    \langle S'\rangle_\mathrm{eq} = \langle D'\rangle_\mathrm{eq} &= 0 \label{eq:1orderIdentities} \\
    \langle D' S'\rangle_\mathrm{eq} &= 0 \label{eq:2orderIdentity1} \\
    \left\langle D''-D'^2-\frac{S'}{4}\right\rangle_\mathrm{eq} &= 0 \\
    \left\langle D'^2 S'-D''S'+\frac{S'^2}{12}\right\rangle_\mathrm{eq} &= 0. \\
    \;\;\vdots \nonumber
\end{align}
Let us calculate the covariance $\langle A;B \rangle \coloneqq \langle AB\rangle - \langle A\rangle \langle B \rangle$ for the first-order term in the response relation for a state observable, Eq.~\eqref{eq:ResponseStateO}.
We find $\langle S';O_t\rangle_\mathrm{eq} = \langle S' O_t\rangle_\mathrm{eq}$ using Eq.~\eqref{eq:1orderIdentities}.
For the second-order response we consider the joint cumulant or connected correlation function of third order, which is defined according to
\begin{align}
\begin{split}
    \langle A;B;C \rangle &\coloneqq \langle ABC\rangle -\langle AB\rangle \langle C \rangle - \langle AC \rangle \langle B \rangle \\
    &\quad- \langle A\rangle \langle BC\rangle + 2 \langle A \rangle \langle B \rangle \langle C \rangle.
    \end{split}
\end{align}
We find $\langle D';S';O_t\rangle_\mathrm{eq} = \langle D'S'O_t\rangle_\mathrm{eq}$ using Eqs.~\eqref{eq:1orderIdentities},\eqref{eq:2orderIdentity1}.
For the first two orders the expressions in terms of connected correlation functions follow trivially, because additional terms vanish individually.
This is different for the third-order terms. 
We find 
\begin{align}
\begin{split}
 \frac{1}{6} \left\langle \left[ 3 D'^2 - 3 D'' + \frac{S'^2}{4} \right] S' O_t  \right\rangle_\mathrm{eq} \quad \quad \quad \quad & \\= \frac{1}{2} \Big[\langle D';D';S';O_t\rangle_\mathrm{eq} - \langle D'';S';O_t\rangle_\mathrm{eq}& \\
 + \frac{1}{12} \langle S';S';S';O_t\rangle_\mathrm{eq}& \Big]  ,
 \end{split}\label{eq:ThirdOrderTermsConnected}
\end{align}
but the three terms cannot be replaced with connected correlation functions individually, i.e.~$\langle D';D';S';O_t\rangle \neq \langle D'^2 S' O_t\rangle_\mathrm{eq}$, $\langle D'';S';O_t\rangle_\mathrm{eq} \neq \langle D'' S' O_t\rangle_\mathrm{eq}$ and $\langle S';S';S';O_t\rangle_\mathrm{eq} \neq \langle S'^3O_t\rangle_\mathrm{eq}$.
Eq.~\eqref{eq:ThirdOrderTermsConnected} contains the connected correlation function of fourth order. 
We do not give its definition explicitly, but provide the general definition of the connected correlation function of order $n$~\cite{peccati_wiener_2011},
\begin{align}
    \langle A_1;\dots;A_n\rangle \coloneqq \sum_\pi (|\pi|-1)!(-1)^{|\pi|-1} \prod_{B \in \pi} \left\langle \prod_{i\in B}A_i \right\rangle    \label{eq:ConnectedCorrelationFunction}
\end{align}
with the sum over all partitions $\{1, \dots, n \}$. 
$|\pi|$ denotes the number of parts in a partition and $B$ refers to all blocks of a partition.

Inserting all relations, we can express Eq.~\eqref{eq:ResponseStateO} in terms of connected correlation functions only and arrive at Eq.~\eqref{eq:ResponseStateOConnectedCorrelations}.
The derivation of a response relation for a general path observable up to second order, Eq.~\eqref{eq:ResponseOConnectedCorrelations}, follows in analogy.

\section{$n-$th order memory kernel}\label{app:Gamman}

In this appendix we set $\dot{x}_t \coloneqq \varepsilon \dot{\Tilde{x}}_t$, with $\varepsilon$ defined as the dimensionless strength of perturbation.
This allows to expand and take derivatives with respect to $\varepsilon$, which is simpler than considering time-dependent velocities.
With this setting the non-equilibrium average of the force can be expressed as
\begin{align}
\begin{split}
    \langle F_t\rangle = \underbrace{\langle F_t\rangle_\mathrm{eq}}_{=0} + \sum_{n=1}^\infty \frac{\varepsilon^n}{n !} \Bigg\langle \frac{\mathrm{d}^n}{\mathrm{d}\varepsilon^n} \Big(e^{-\varepsilon D'(\omega)+ \frac{\varepsilon}{2} S'(\omega)- \frac{\varepsilon^2}{2}D''(\omega)}\quad&\\
    - e^{-\varepsilon D'(\omega)-\frac{\varepsilon}{2}S'(\omega) - \frac{\varepsilon^2}{2} D''(\omega)} \Big)\Bigg|_{\varepsilon=0} F_t \Bigg\rangle_\mathrm{eq}&
    \end{split}
\end{align}
by expanding the perturbed action $\mathcal{A}_\varepsilon $ in a Taylor series in $\varepsilon$, $\mathcal{A}_\varepsilon = \varepsilon D' -\varepsilon S'/2+\varepsilon^2 D''/2$~\cite{holsten_thermodynamic_2021}.

Applying general Leibniz rule on the $n-$th derivative of the exponential we can write~\cite{holsten_thermodynamic_2021}
\begin{align}
\begin{split}
    \langle F_t\rangle &= \sum_{n=1}^\infty \frac{\varepsilon^n}{n !} \Bigg\langle \sum_{i=0}^n \binom{n}{i} \frac{\mathrm{d}^{n-i}}{\mathrm{d}\varepsilon^{n-i}} e^{-\varepsilon D'(\omega)-\frac{\varepsilon^2}{2} D''(\omega)}\Bigg|_{\varepsilon=0} \\
    &\quad\times\left[ \left(\frac{S'(\omega)}{2} \right)^i - \left(-\frac{S'(\omega)}{2} \right)^i \right] F_t \Bigg\rangle_\mathrm{eq}.
    \end{split}
    \label{eq:ForceResponseInfty}
\end{align}
Let us first consider the NT part.
From Fa\`a di Bruno's formula expressed in terms of complete Bell polynomials $B_n$, $\frac{\mathrm{d}^n}{\mathrm{d}x^n} e^{f(x)} = e^{f(x)} B_n (f'(x),f''(x),\dots,f^{(n)}(x))$~\cite{bruno_sullo_1855} we find
\begin{align}
\begin{split}
    \frac{\mathrm{d}^{n-i}}{\mathrm{d}\varepsilon^{n-i}} &e^{-\varepsilon D'(\omega)-\frac{\varepsilon^2}{2} D''(\omega)}\Bigg|_{\varepsilon=0} \\
    &\quad\quad\quad=  B_{n-i} \left(-D',-D'',0,\dots,0\right).
    \end{split}
    \label{eq:BellpolynomialsFrenetic}
\end{align}
Eq.~\eqref{eq:BellpolynomialsFrenetic} thus contains $n-i$ integrals which can be taken out of the Bell polynomials when functional derivatives are applied,
\begin{widetext}
\begin{align}
    \frac{\mathrm{d}^{n-i}}{\mathrm{d}\varepsilon^{n-i}} e^{-D(\omega)}\Bigg|_{\varepsilon=0}  = \frac{1}{(n-i)!}\int_{t_0}^t \mathrm{d}s_1\cdots  \int_{t_0}^t \mathrm{d}s_{n-i} \, \dot{x}_{s_1}\cdots \dot{x}_{s_{n-i}} \frac{\delta^{n-i}}{\delta \dot{x}_{s_1}\cdots \delta \dot{x}_{s_{n-i}}} B_{n-i} \left(-D',-D'',0,\dots,0\right). 
    \label{eq:BellpolynomialsFreneticFunctionalDerivatives}
\end{align}
The entropic part takes on the form
\begin{align}
    \left( \frac{S'(\omega)}{2} \right)^i - \left(-\frac{S'(\omega)}{2} \right)^i =  \left( \frac{\beta}{2} \right)^i \left( 1-(-1)^i \right) \int_{t_0}^t \mathrm{d}s_1 \cdots \int_{t_0}^t \mathrm{d}s_i \, \dot{x}_{s_1} \cdots \dot{x}_{s_i} F_{s_1} \cdots F_{s_i} .
    \label{eq:EntropicPartIntegrals}
\end{align}
Inserting Eqs.~\eqref{eq:BellpolynomialsFreneticFunctionalDerivatives} and \eqref{eq:EntropicPartIntegrals} back into Eq.~\eqref{eq:ForceResponseInfty} we obtain
\begin{align}
\begin{split}
    \langle F_t\rangle &= \sum_{n=1}^\infty \frac{\varepsilon^n}{n!} \int_{t_0}^t \mathrm{d}s_1 \cdots \int_{t_0}^t \mathrm{d}s_n \, \dot{x}_{s_1} \cdots \dot{x}_{s_n} \sum_{i=1}^n \frac{\left(1-(-1)^i \right)}{(n-i)!} \binom{n}{i} \left( \frac{\beta}{2}\right)^i \\
    &\quad\times\left\langle  \frac{\delta^{n-i}}{\delta \dot{x}_{t-s_1}\cdots \delta \dot{x}_{t-s_{n-i}}} B_{n-i} \left(D',-D'',0,\dots,0\right) F_{t-s_{n-i+1}} \cdots F_{t-s_n} F_0 \right\rangle_\mathrm{eq},
    \end{split}
\end{align}
where we have shifted all time arguments by time $t$ and applied time-reversal (resulting in a change of sign of NT components of even order) to follow the notation in Eqs.~\eqref{eq:MeanForce}-\eqref{eq:ForceCum3}.
Applying Eq.~\eqref{eq:Gammamn}, the $n-$th order memory kernel for the mean force can thus be expressed as
\begin{align}
\begin{split}
    \Gamma_{s_1,\dots,s_n}^{(0,n)} = &\frac{\beta}{(n !)^2} \sum_{\pi\in S_n} \sum_{i=1}^n \frac{\left(1-(-1)^i\right)}{(n-i)!} \binom{n}{i} \left(\frac{\beta}{2}\right)^i \\
    &\times\left\langle \frac{\delta^{n-i}}{\delta \dot{x}_{s_{\pi(1)}} \cdots \delta \dot{x}_{s_{\pi(n-i)}}} B_{n-i}\left(D',-D'',0,\dots , 0\right) F_{s_{\pi(n-i+1)}} \cdots F_{s_{\pi(n)}}  F_0\right\rangle_\mathrm{eq}.
    \label{eq:Gamman}
    \end{split}
\end{align}
\end{widetext}
Note that the notion of connected correlation functions is not included in the above derivation.
Eq.~\eqref{eq:Gamman} expresses $\Gamma^{(0,n)}$ in terms of equilibrium correlation functions only.

\section{Convergence of connected correlation functions}\label{app:ConvergenceConnectedCorr}
\begin{figure}[h]
    \includegraphics{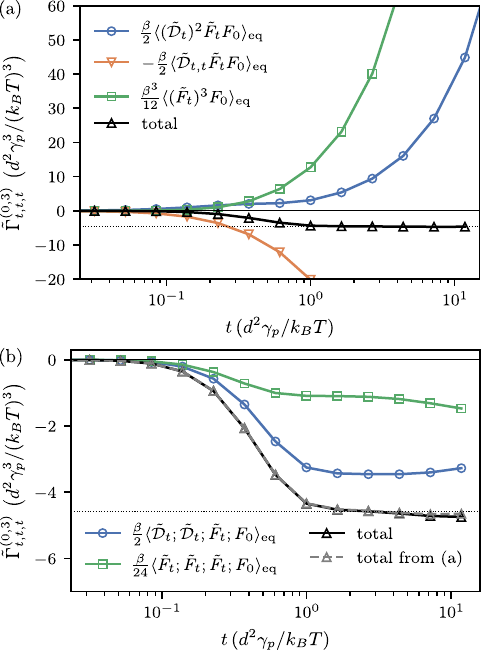}
    \caption{Running integral of the third-order memory kernel $\tilde{\Gamma}^{(0,3)}_{t,t,t} = \int_0^t \mathrm{d}s \int_0^t \mathrm{d}s' \int_0^t \mathrm{d}s''\, \Gamma_{s,s',s''}^{(0,3)}$ calculated using (a) the (standard) response relation, Eq.~\eqref{eq:ResponseStateO}, and (b) the response relation in terms of connected correlation functions, Eq.~\eqref{eq:ResponseStateOConnectedCorrelations}. Colored curves show individual terms and black curves their total. The dotted lines are identical in (a) and (b) and indicate the limit $\lim_{t \to \infty} \tilde{\Gamma}_{t,t,t}^{(0,3)}$ used for further analysis. The black line from (a) is added as dashed gray line to (b).}
    \label{fig:thirdorderforceresponse}
\end{figure}

How can we obtain the time-integrated kernel $\hat{\Gamma}^{(0,3)}_{z=0,z=0,z=0}$ given by Eq.~\eqref{eq:Gamma3general} efficiently?
We first calculate running integrals $\tilde{A}_t \coloneqq \int_0^t \mathrm{d}s\, A_s$ and afterwards compute respective (connected) correlation functions as function of time $t$.
We then check convergence and read off the long-time limit.
First, we follow this approach to naively compute the running integral of $\Gamma^{(0,3)}$ over all times,  $\tilde{\Gamma}^{(0,3)}_{t,t,t} = \int_0^t \mathrm{d}s \int_0^t \mathrm{d}s' \int_0^t \mathrm{d}s'' \, \Gamma_{s,s',s''}^{(0,3)}$, via the response relation in terms of correlation functions, Eq.~\eqref{eq:ResponseStateO}, i.e.~Eq.~\eqref{eq:Gamman} in Appendix~\ref{app:Gamman}.
The result for the scenario of direct driving is shown in Fig.~\ref{fig:thirdorderforceresponse}(a).
We observe convergence to $\hat{\Gamma}^{(0,3)}_{z=0,z=0,z=0} \approx -4.5\, d^2 \gamma_p^3/(k_B T)^3$ for the total (black line), however the three individual terms diverge (colored lines).
In contrast, when applying the response relation in terms of connected correlation functions, Eq.~\eqref{eq:ResponseStateOConnectedCorrelations}, we observe convergence of the individual terms as well, see Fig.~\ref{fig:thirdorderforceresponse}(b). 
The curves for Fig.~\ref{fig:thirdorderforceresponse}(a) and (b) were obtained in the same simulation, i.e.~with identical statistics.
The two total curves (black and gray lines) are almost identical and converge to the same limiting value. 
Connected correlation functions therefore have an important interpretation, but little influence on the convergence of the total response.

\section{Transient force covariance for the scenario of indirect driving}\label{app:OscillationsIndirectDriving}

\begin{figure}[h]
    \centering
    \includegraphics{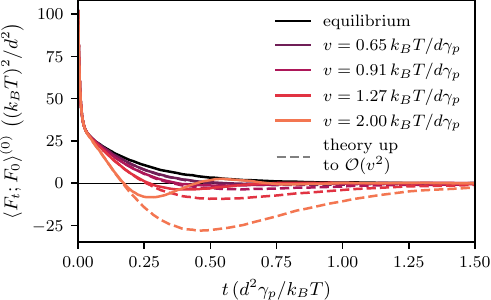}
    \caption{Transient force covariance for different driving velocities $v$ (colored lines) for the scenario of indirect driving with $\kappa = 100\, k_B T/d^2$. Dashed lines show response curves according to Eq.~\eqref{eq:ForceCov} up to $\mathcal{O}(v^2)$.}
    \label{fig:TransientForceCovKappa100}
\end{figure}

Fig.~\ref{fig:TransientForceCovKappa100} shows simulations of the transient force covariance for different driving velocities $v$ (solid lines) for the scenario of indirect driving with $\kappa = 100\, k_B T/d^2$.
The dashed lines correspond to theoretical response curves up to second order according to Eq.~\eqref{eq:ForceCov}.
At time zero, we find $\langle F_0;F_0\rangle^{(0)} = \kappa k_B T$ (compare Eq.~\eqref{eq:ForceVarianceEqIndirect}) which is followed by a fast decay $\propto e^{-\frac{\kappa}{\gamma_p}t}$ with the relaxation time of the bare probe in the trap. 
For larger driving velocity the force covariance exhibits clear oscillations in agreement with experiments in micellar systems~\cite{berner_oscillating_2018,jain_two_2021}.
This phenomenon already appears in the second-order response (dashed lines).

\section*{References}

\begin{thebibliography}{67}%
\makeatletter
\providecommand \@ifxundefined [1]{%
 \@ifx{#1\undefined}
}%
\providecommand \@ifnum [1]{%
 \ifnum #1\expandafter \@firstoftwo
 \else \expandafter \@secondoftwo
 \fi
}%
\providecommand \@ifx [1]{%
 \ifx #1\expandafter \@firstoftwo
 \else \expandafter \@secondoftwo
 \fi
}%
\providecommand \natexlab [1]{#1}%
\providecommand \enquote  [1]{``#1''}%
\providecommand \bibnamefont  [1]{#1}%
\providecommand \bibfnamefont [1]{#1}%
\providecommand \citenamefont [1]{#1}%
\providecommand \href@noop [0]{\@secondoftwo}%
\providecommand \href [0]{\begingroup \@sanitize@url \@href}%
\providecommand \@href[1]{\@@startlink{#1}\@@href}%
\providecommand \@@href[1]{\endgroup#1\@@endlink}%
\providecommand \@sanitize@url [0]{\catcode `\\12\catcode `\$12\catcode
  `\&12\catcode `\#12\catcode `\^12\catcode `\_12\catcode `\%12\relax}%
\providecommand \@@startlink[1]{}%
\providecommand \@@endlink[0]{}%
\providecommand \url  [0]{\begingroup\@sanitize@url \@url }%
\providecommand \@url [1]{\endgroup\@href {#1}{\urlprefix }}%
\providecommand \urlprefix  [0]{URL }%
\providecommand \Eprint [0]{\href }%
\providecommand \doibase [0]{https://doi.org/}%
\providecommand \bibinfo  [0]{\@secondoftwo}%
\providecommand \bibfield  [0]{\@secondoftwo}%
\providecommand \translation [1]{[#1]}%
\providecommand \BibitemOpen [0]{}%
\providecommand \bibitemStop [0]{}%
\providecommand \bibitemNoStop [0]{.\EOS\space}%
\providecommand \EOS [0]{\spacefactor3000\relax}%
\providecommand \BibitemShut  [1]{\csname bibitem#1\endcsname}%
\let\auto@bib@innerbib\@empty
\bibitem [{\citenamefont {Dhont}(1996)}]{dhont_introduction_1996}%
  \BibitemOpen
  \bibfield  {author} {\bibinfo {author} {\bibfnamefont {J.}~\bibnamefont
  {Dhont}},\ }\href {https://books.google.at/books?id=mmArTF5SJ9oC} {\emph
  {\bibinfo {title} {An {Introduction} to {Dynamics} of {Colloids}}}}\
  (\bibinfo  {publisher} {Elsevier Science},\ \bibinfo {year}
  {1996})\BibitemShut {NoStop}%
\bibitem [{\citenamefont {Baiesi}, \citenamefont {Iubini},\ and\ \citenamefont
  {Orlandini}(2021)}]{baiesi_rise_2021}%
  \BibitemOpen
  \bibfield  {author} {\bibinfo {author} {\bibfnamefont {M.}~\bibnamefont
  {Baiesi}}, \bibinfo {author} {\bibfnamefont {S.}~\bibnamefont {Iubini}},\
  and\ \bibinfo {author} {\bibfnamefont {E.}~\bibnamefont {Orlandini}},\
  }\bibfield  {title} {\enquote {\bibinfo {title} {The rise and fall of
  branching: {A} slowing down mechanism in relaxing wormlike micellar
  networks},}\ }\href {https://doi.org/10.1063/5.0072374} {\bibfield  {journal}
  {\bibinfo  {journal} {The Journal of Chemical Physics}\ }\textbf {\bibinfo
  {volume} {155}},\ \bibinfo {pages} {214905} (\bibinfo {year}
  {2021})}\BibitemShut {NoStop}%
\bibitem [{\citenamefont {Ginot}\ \emph {et~al.}(2022)\citenamefont {Ginot},
  \citenamefont {Caspers}, \citenamefont {Reinalter}, \citenamefont {Kumar},
  \citenamefont {Krüger},\ and\ \citenamefont
  {Bechinger}}]{ginot_recoil_2022}%
  \BibitemOpen
  \bibfield  {author} {\bibinfo {author} {\bibfnamefont {F.}~\bibnamefont
  {Ginot}}, \bibinfo {author} {\bibfnamefont {J.}~\bibnamefont {Caspers}},
  \bibinfo {author} {\bibfnamefont {L.~F.}\ \bibnamefont {Reinalter}}, \bibinfo
  {author} {\bibfnamefont {K.~K.}\ \bibnamefont {Kumar}}, \bibinfo {author}
  {\bibfnamefont {M.}~\bibnamefont {Krüger}},\ and\ \bibinfo {author}
  {\bibfnamefont {C.}~\bibnamefont {Bechinger}},\ }\bibfield  {title} {\enquote
  {\bibinfo {title} {Recoil experiments determine the eigenmodes of
  viscoelastic fluids},}\ }\href {https://doi.org/10.1088/1367-2630/aca8c7}
  {\bibfield  {journal} {\bibinfo  {journal} {New J. Phys.}\ }\textbf {\bibinfo
  {volume} {24}},\ \bibinfo {pages} {123013} (\bibinfo {year}
  {2022})}\BibitemShut {NoStop}%
\bibitem [{\citenamefont {Zwanzig}(1961)}]{zwanzig_memory_1961}%
  \BibitemOpen
  \bibfield  {author} {\bibinfo {author} {\bibfnamefont {R.}~\bibnamefont
  {Zwanzig}},\ }\bibfield  {title} {\enquote {\bibinfo {title} {Memory
  {Effects} in {Irreversible} {Thermodynamics}},}\ }\href
  {https://doi.org/10.1103/PhysRev.124.983} {\bibfield  {journal} {\bibinfo
  {journal} {Phys. Rev.}\ }\textbf {\bibinfo {volume} {124}},\ \bibinfo {pages}
  {983--992} (\bibinfo {year} {1961})}\BibitemShut {NoStop}%
\bibitem [{\citenamefont {Mori}(1965)}]{mori_transport_1965}%
  \BibitemOpen
  \bibfield  {author} {\bibinfo {author} {\bibfnamefont {H.}~\bibnamefont
  {Mori}},\ }\bibfield  {title} {\enquote {\bibinfo {title} {Transport,
  {Collective} {Motion}, and {Brownian} {Motion}},}\ }\href
  {https://doi.org/10.1143/PTP.33.423} {\bibfield  {journal} {\bibinfo
  {journal} {Progress of Theoretical Physics}\ }\textbf {\bibinfo {volume}
  {33}},\ \bibinfo {pages} {423--455} (\bibinfo {year} {1965})}\BibitemShut
  {NoStop}%
\bibitem [{\citenamefont {Zwanzig}(2001)}]{zwanzig_nonequilibrium_2001}%
  \BibitemOpen
  \bibfield  {author} {\bibinfo {author} {\bibfnamefont {R.}~\bibnamefont
  {Zwanzig}},\ }\href {https://books.google.de/books?id=4cI5136OdoMC} {\emph
  {\bibinfo {title} {Nonequilibrium {Statistical} {Mechanics}}}}\ (\bibinfo
  {publisher} {Oxford University Press},\ \bibinfo {year} {2001})\BibitemShut
  {NoStop}%
\bibitem [{\citenamefont {Grabert}(1982)}]{grabert_projection_1982}%
  \BibitemOpen
  \bibfield  {author} {\bibinfo {author} {\bibfnamefont {H.}~\bibnamefont
  {Grabert}},\ }\href {https://doi.org/10.1007/BFb0044591} {\emph {\bibinfo
  {title} {Projection {Operator} {Techniques} in {Nonequilibrium} {Statistical}
  {Mechanics}}}},\ \bibinfo {series} {Springer {Tracts} in {Modern} {Physics}},
  Vol.~\bibinfo {volume} {95}\ (\bibinfo  {publisher} {Springer},\ \bibinfo
  {address} {Berlin, Heidelberg},\ \bibinfo {year} {1982})\BibitemShut
  {NoStop}%
\bibitem [{\citenamefont {Caldeira}\ and\ \citenamefont
  {Leggett}(1981)}]{caldeira_influence_1981}%
  \BibitemOpen
  \bibfield  {author} {\bibinfo {author} {\bibfnamefont {A.~O.}\ \bibnamefont
  {Caldeira}}\ and\ \bibinfo {author} {\bibfnamefont {A.~J.}\ \bibnamefont
  {Leggett}},\ }\bibfield  {title} {\enquote {\bibinfo {title} {Influence of
  {Dissipation} on {Quantum} {Tunneling} in {Macroscopic} {Systems}},}\ }\href
  {https://doi.org/10.1103/PhysRevLett.46.211} {\bibfield  {journal} {\bibinfo
  {journal} {Phys. Rev. Lett.}\ }\textbf {\bibinfo {volume} {46}},\ \bibinfo
  {pages} {211--214} (\bibinfo {year} {1981})}\BibitemShut {NoStop}%
\bibitem [{\citenamefont {Götze}(2008)}]{gotze_complex_2008}%
  \BibitemOpen
  \bibfield  {author} {\bibinfo {author} {\bibfnamefont {W.}~\bibnamefont
  {Götze}},\ }\href
  {https://doi.org/10.1093/acprof:oso/9780199235346.001.0001} {\emph {\bibinfo
  {title} {Complex {Dynamics} of {Glass}-{Forming} {Liquids}: {A}
  {Mode}-{Coupling} {Theory}}}}\ (\bibinfo  {publisher} {Oxford University
  Press},\ \bibinfo {year} {2008})\BibitemShut {NoStop}%
\bibitem [{\citenamefont {Janssen}(2018)}]{janssen_mode-coupling_2018}%
  \BibitemOpen
  \bibfield  {author} {\bibinfo {author} {\bibfnamefont {L.~M.~C.}\
  \bibnamefont {Janssen}},\ }\bibfield  {title} {\enquote {\bibinfo {title}
  {Mode-{Coupling} {Theory} of the {Glass} {Transition}: {A} {Primer}},}\
  }\href {https://www.frontiersin.org/articles/10.3389/fphy.2018.00097}
  {\bibfield  {journal} {\bibinfo  {journal} {Frontiers in Physics}\ }\textbf
  {\bibinfo {volume} {6}} (\bibinfo {year} {2018})}\BibitemShut {NoStop}%
\bibitem [{\citenamefont {Fuchs}, \citenamefont {Götze},\ and\ \citenamefont
  {Mayr}(1998)}]{fuchs_asymptotic_1998}%
  \BibitemOpen
  \bibfield  {author} {\bibinfo {author} {\bibfnamefont {M.}~\bibnamefont
  {Fuchs}}, \bibinfo {author} {\bibfnamefont {W.}~\bibnamefont {Götze}},\ and\
  \bibinfo {author} {\bibfnamefont {M.~R.}\ \bibnamefont {Mayr}},\ }\bibfield
  {title} {\enquote {\bibinfo {title} {Asymptotic laws for tagged-particle
  motion in glassy systems},}\ }\href
  {https://doi.org/10.1103/PhysRevE.58.3384} {\bibfield  {journal} {\bibinfo
  {journal} {Phys. Rev. E}\ }\textbf {\bibinfo {volume} {58}},\ \bibinfo
  {pages} {3384--3399} (\bibinfo {year} {1998})}\BibitemShut {NoStop}%
\bibitem [{\citenamefont {van Zanten}\ and\ \citenamefont
  {Rufener}(2000)}]{van_zanten_brownian_2000}%
  \BibitemOpen
  \bibfield  {author} {\bibinfo {author} {\bibfnamefont {J.~H.}\ \bibnamefont
  {van Zanten}}\ and\ \bibinfo {author} {\bibfnamefont {K.~P.}\ \bibnamefont
  {Rufener}},\ }\bibfield  {title} {\enquote {\bibinfo {title} {Brownian motion
  in a single relaxation time {Maxwell} fluid},}\ }\href
  {https://doi.org/10.1103/PhysRevE.62.5389} {\bibfield  {journal} {\bibinfo
  {journal} {Phys. Rev. E}\ }\textbf {\bibinfo {volume} {62}},\ \bibinfo
  {pages} {5389--5396} (\bibinfo {year} {2000})}\BibitemShut {NoStop}%
\bibitem [{\citenamefont {Lu}\ and\ \citenamefont
  {Solomon}(2002)}]{lu_probe_2002}%
  \BibitemOpen
  \bibfield  {author} {\bibinfo {author} {\bibfnamefont {Q.}~\bibnamefont
  {Lu}}\ and\ \bibinfo {author} {\bibfnamefont {M.~J.}\ \bibnamefont
  {Solomon}},\ }\bibfield  {title} {\enquote {\bibinfo {title} {Probe size
  effects on the microrheology of associating polymer solutions},}\ }\href
  {https://doi.org/10.1103/PhysRevE.66.061504} {\bibfield  {journal} {\bibinfo
  {journal} {Phys. Rev. E}\ }\textbf {\bibinfo {volume} {66}},\ \bibinfo
  {pages} {061504} (\bibinfo {year} {2002})}\BibitemShut {NoStop}%
\bibitem [{\citenamefont {Van Der~Gucht}\ \emph {et~al.}(2003)\citenamefont
  {Van Der~Gucht}, \citenamefont {Besseling}, \citenamefont {Knoben},
  \citenamefont {Bouteiller},\ and\ \citenamefont
  {Cohen~Stuart}}]{van_der_gucht_brownian_2003}%
  \BibitemOpen
  \bibfield  {author} {\bibinfo {author} {\bibfnamefont {J.}~\bibnamefont {Van
  Der~Gucht}}, \bibinfo {author} {\bibfnamefont {N.~A.~M.}\ \bibnamefont
  {Besseling}}, \bibinfo {author} {\bibfnamefont {W.}~\bibnamefont {Knoben}},
  \bibinfo {author} {\bibfnamefont {L.}~\bibnamefont {Bouteiller}},\ and\
  \bibinfo {author} {\bibfnamefont {M.~A.}\ \bibnamefont {Cohen~Stuart}},\
  }\bibfield  {title} {\enquote {\bibinfo {title} {Brownian particles in
  supramolecular polymer solutions},}\ }\href
  {https://doi.org/10.1103/PhysRevE.67.051106} {\bibfield  {journal} {\bibinfo
  {journal} {Phys. Rev. E}\ }\textbf {\bibinfo {volume} {67}},\ \bibinfo
  {pages} {051106} (\bibinfo {year} {2003})}\BibitemShut {NoStop}%
\bibitem [{\citenamefont {Caspers}\ \emph {et~al.}(2023)\citenamefont
  {Caspers}, \citenamefont {Ditz}, \citenamefont {Krishna~Kumar}, \citenamefont
  {Ginot}, \citenamefont {Bechinger}, \citenamefont {Fuchs},\ and\
  \citenamefont {Krüger}}]{caspers_how_2023}%
  \BibitemOpen
  \bibfield  {author} {\bibinfo {author} {\bibfnamefont {J.}~\bibnamefont
  {Caspers}}, \bibinfo {author} {\bibfnamefont {N.}~\bibnamefont {Ditz}},
  \bibinfo {author} {\bibfnamefont {K.}~\bibnamefont {Krishna~Kumar}}, \bibinfo
  {author} {\bibfnamefont {F.}~\bibnamefont {Ginot}}, \bibinfo {author}
  {\bibfnamefont {C.}~\bibnamefont {Bechinger}}, \bibinfo {author}
  {\bibfnamefont {M.}~\bibnamefont {Fuchs}},\ and\ \bibinfo {author}
  {\bibfnamefont {M.}~\bibnamefont {Krüger}},\ }\bibfield  {title} {\enquote
  {\bibinfo {title} {How are mobility and friction related in viscoelastic
  fluids?}}\ }\href {https://doi.org/10.1063/5.0129639} {\bibfield  {journal}
  {\bibinfo  {journal} {J. Chem. Phys.}\ }\textbf {\bibinfo {volume} {158}},\
  \bibinfo {pages} {024901} (\bibinfo {year} {2023})}\BibitemShut {NoStop}%
\bibitem [{\citenamefont {Doerries}, \citenamefont {Loos},\ and\ \citenamefont
  {Klapp}(2021)}]{doerries_correlation_2021}%
  \BibitemOpen
  \bibfield  {author} {\bibinfo {author} {\bibfnamefont {T.~J.}\ \bibnamefont
  {Doerries}}, \bibinfo {author} {\bibfnamefont {S.~A.~M.}\ \bibnamefont
  {Loos}},\ and\ \bibinfo {author} {\bibfnamefont {S.~H.~L.}\ \bibnamefont
  {Klapp}},\ }\bibfield  {title} {{ {\bibinfo
  {title} {Correlation functions of non-{Markovian} systems out of equilibrium:
  analytical expressions beyond single-exponential memory},}\ }}\href
  {https://doi.org/10.1088/1742-5468/abdead} {\bibfield  {journal} {\bibinfo
  {journal} {J. Stat. Mech.}\ }\textbf {\bibinfo {volume} {2021}},\ \bibinfo
  {pages} {033202} (\bibinfo {year} {2021})}\BibitemShut {NoStop}%
\bibitem [{\citenamefont {Venturelli}\ and\ \citenamefont
  {Gambassi}(2023)}]{venturelli_memory-induced_2023}%
  \BibitemOpen
  \bibfield  {author} {\bibinfo {author} {\bibfnamefont {D.}~\bibnamefont
  {Venturelli}}\ and\ \bibinfo {author} {\bibfnamefont {A.}~\bibnamefont
  {Gambassi}},\ }\bibfield  {title} {\enquote {\bibinfo {title} {Memory-induced
  oscillations of a driven particle in a dissipative correlated medium},}\
  }\href {https://doi.org/10.1088/1367-2630/acf240} {\bibfield  {journal}
  {\bibinfo  {journal} {New J. Phys.}\ }\textbf {\bibinfo {volume} {25}},\
  \bibinfo {pages} {093025} (\bibinfo {year} {2023})}\BibitemShut {NoStop}%
\bibitem [{\citenamefont {Squires}\ and\ \citenamefont
  {Brady}(2005)}]{squires_simple_2005}%
  \BibitemOpen
  \bibfield  {author} {\bibinfo {author} {\bibfnamefont {T.~M.}\ \bibnamefont
  {Squires}}\ and\ \bibinfo {author} {\bibfnamefont {J.~F.}\ \bibnamefont
  {Brady}},\ }\bibfield  {title} {\enquote {\bibinfo {title} {A simple paradigm
  for active and nonlinear microrheology},}\ }\href
  {https://doi.org/10.1063/1.1960607} {\bibfield  {journal} {\bibinfo
  {journal} {Physics of Fluids}\ }\textbf {\bibinfo {volume} {17}},\ \bibinfo
  {pages} {073101} (\bibinfo {year} {2005})}\BibitemShut {NoStop}%
\bibitem [{\citenamefont {Gazuz}\ \emph {et~al.}(2009)\citenamefont {Gazuz},
  \citenamefont {Puertas}, \citenamefont {Voigtmann},\ and\ \citenamefont
  {Fuchs}}]{gazuz_active_2009}%
  \BibitemOpen
  \bibfield  {author} {\bibinfo {author} {\bibfnamefont {I.}~\bibnamefont
  {Gazuz}}, \bibinfo {author} {\bibfnamefont {A.~M.}\ \bibnamefont {Puertas}},
  \bibinfo {author} {\bibfnamefont {T.}~\bibnamefont {Voigtmann}},\ and\
  \bibinfo {author} {\bibfnamefont {M.}~\bibnamefont {Fuchs}},\ }\bibfield
  {title} {\enquote {\bibinfo {title} {Active and {Nonlinear} {Microrheology}
  in {Dense} {Colloidal} {Suspensions}},}\ }\href
  {https://doi.org/10.1103/PhysRevLett.102.248302} {\bibfield  {journal}
  {\bibinfo  {journal} {Phys. Rev. Lett.}\ }\textbf {\bibinfo {volume} {102}},\
  \bibinfo {pages} {248302} (\bibinfo {year} {2009})}\BibitemShut {NoStop}%
\bibitem [{\citenamefont {Harrer}\ \emph {et~al.}(2012)\citenamefont {Harrer},
  \citenamefont {Winter}, \citenamefont {Horbach}, \citenamefont {Fuchs},\ and\
  \citenamefont {Voigtmann}}]{harrer_force-induced_2012}%
  \BibitemOpen
  \bibfield  {author} {\bibinfo {author} {\bibfnamefont {C.~J.}\ \bibnamefont
  {Harrer}}, \bibinfo {author} {\bibfnamefont {D.}~\bibnamefont {Winter}},
  \bibinfo {author} {\bibfnamefont {J.}~\bibnamefont {Horbach}}, \bibinfo
  {author} {\bibfnamefont {M.}~\bibnamefont {Fuchs}},\ and\ \bibinfo {author}
  {\bibfnamefont {T.}~\bibnamefont {Voigtmann}},\ }\bibfield  {title} {\enquote
  {\bibinfo {title} {Force-induced diffusion in microrheology},}\ }\href
  {https://doi.org/10.1088/0953-8984/24/46/464105} {\bibfield  {journal}
  {\bibinfo  {journal} {J. Phys.: Condens. Matter}\ }\textbf {\bibinfo {volume}
  {24}},\ \bibinfo {pages} {464105} (\bibinfo {year} {2012})}\BibitemShut
  {NoStop}%
\bibitem [{\citenamefont {Jayaraman}\ and\ \citenamefont
  {Belmonte}(2003)}]{jayaraman_oscillations_2003}%
  \BibitemOpen
  \bibfield  {author} {\bibinfo {author} {\bibfnamefont {A.}~\bibnamefont
  {Jayaraman}}\ and\ \bibinfo {author} {\bibfnamefont {A.}~\bibnamefont
  {Belmonte}},\ }\bibfield  {title} {\enquote {\bibinfo {title} {Oscillations
  of a solid sphere falling through a wormlike micellar fluid},}\ }\href
  {https://doi.org/10.1103/PhysRevE.67.065301} {\bibfield  {journal} {\bibinfo
  {journal} {Phys. Rev. E}\ }\textbf {\bibinfo {volume} {67}},\ \bibinfo
  {pages} {065301} (\bibinfo {year} {2003})}\BibitemShut {NoStop}%
\bibitem [{\citenamefont {Handzy}\ and\ \citenamefont
  {Belmonte}(2004)}]{handzy_oscillatory_2004}%
  \BibitemOpen
  \bibfield  {author} {\bibinfo {author} {\bibfnamefont {N.~Z.}\ \bibnamefont
  {Handzy}}\ and\ \bibinfo {author} {\bibfnamefont {A.}~\bibnamefont
  {Belmonte}},\ }\bibfield  {title} {\enquote {\bibinfo {title} {Oscillatory
  {Rise} of {Bubbles} in {Wormlike} {Micellar} {Fluids} with {Different}
  {Microstructures}},}\ }\href {https://doi.org/10.1103/PhysRevLett.92.124501}
  {\bibfield  {journal} {\bibinfo  {journal} {Phys. Rev. Lett.}\ }\textbf
  {\bibinfo {volume} {92}},\ \bibinfo {pages} {124501} (\bibinfo {year}
  {2004})}\BibitemShut {NoStop}%
\bibitem [{\citenamefont {Berner}\ \emph {et~al.}(2018)\citenamefont {Berner},
  \citenamefont {Müller}, \citenamefont {Gomez-Solano}, \citenamefont
  {Krüger},\ and\ \citenamefont {Bechinger}}]{berner_oscillating_2018}%
  \BibitemOpen
  \bibfield  {author} {\bibinfo {author} {\bibfnamefont {J.}~\bibnamefont
  {Berner}}, \bibinfo {author} {\bibfnamefont {B.}~\bibnamefont {Müller}},
  \bibinfo {author} {\bibfnamefont {J.~R.}\ \bibnamefont {Gomez-Solano}},
  \bibinfo {author} {\bibfnamefont {M.}~\bibnamefont {Krüger}},\ and\ \bibinfo
  {author} {\bibfnamefont {C.}~\bibnamefont {Bechinger}},\ }\bibfield  {title}
  {\enquote {\bibinfo {title} {Oscillating modes of driven colloids in
  overdamped systems},}\ }\href {https://doi.org/10.1038/s41467-018-03345-2}
  {\bibfield  {journal} {\bibinfo  {journal} {Nat Commun}\ }\textbf {\bibinfo
  {volume} {9}},\ \bibinfo {pages} {999} (\bibinfo {year} {2018})}\BibitemShut
  {NoStop}%
\bibitem [{\citenamefont {Jain}\ \emph {et~al.}(2021)\citenamefont {Jain},
  \citenamefont {Ginot}, \citenamefont {Berner}, \citenamefont {Bechinger},\
  and\ \citenamefont {Krüger}}]{jain_two_2021}%
  \BibitemOpen
  \bibfield  {author} {\bibinfo {author} {\bibfnamefont {R.}~\bibnamefont
  {Jain}}, \bibinfo {author} {\bibfnamefont {F.}~\bibnamefont {Ginot}},
  \bibinfo {author} {\bibfnamefont {J.}~\bibnamefont {Berner}}, \bibinfo
  {author} {\bibfnamefont {C.}~\bibnamefont {Bechinger}},\ and\ \bibinfo
  {author} {\bibfnamefont {M.}~\bibnamefont {Krüger}},\ }\bibfield  {title}
  {\enquote {\bibinfo {title} {Two step micro-rheological behavior in a
  viscoelastic fluid},}\ }\href {https://doi.org/10.1063/5.0048320} {\bibfield
  {journal} {\bibinfo  {journal} {J. Chem. Phys.}\ }\textbf {\bibinfo {volume}
  {154}},\ \bibinfo {pages} {184904} (\bibinfo {year} {2021})}\BibitemShut
  {NoStop}%
\bibitem [{\citenamefont {Winter}\ \emph {et~al.}(2012)\citenamefont {Winter},
  \citenamefont {Horbach}, \citenamefont {Virnau},\ and\ \citenamefont
  {Binder}}]{winter_active_2012}%
  \BibitemOpen
  \bibfield  {author} {\bibinfo {author} {\bibfnamefont {D.}~\bibnamefont
  {Winter}}, \bibinfo {author} {\bibfnamefont {J.}~\bibnamefont {Horbach}},
  \bibinfo {author} {\bibfnamefont {P.}~\bibnamefont {Virnau}},\ and\ \bibinfo
  {author} {\bibfnamefont {K.}~\bibnamefont {Binder}},\ }\bibfield  {title}
  {\enquote {\bibinfo {title} {Active {Nonlinear} {Microrheology} in a
  {Glass}-{Forming} {Yukawa} {Fluid}},}\ }\href
  {https://doi.org/10.1103/PhysRevLett.108.028303} {\bibfield  {journal}
  {\bibinfo  {journal} {Phys. Rev. Lett.}\ }\textbf {\bibinfo {volume} {108}},\
  \bibinfo {pages} {028303} (\bibinfo {year} {2012})}\BibitemShut {NoStop}%
\bibitem [{\citenamefont {Bénichou}\ \emph {et~al.}(2013)\citenamefont
  {Bénichou}, \citenamefont {Bodrova}, \citenamefont {Chakraborty},
  \citenamefont {Illien}, \citenamefont {Law}, \citenamefont
  {Mejía-Monasterio}, \citenamefont {Oshanin},\ and\ \citenamefont
  {Voituriez}}]{benichou_geometry-induced_2013}%
  \BibitemOpen
  \bibfield  {author} {\bibinfo {author} {\bibfnamefont {O.}~\bibnamefont
  {Bénichou}}, \bibinfo {author} {\bibfnamefont {A.}~\bibnamefont {Bodrova}},
  \bibinfo {author} {\bibfnamefont {D.}~\bibnamefont {Chakraborty}}, \bibinfo
  {author} {\bibfnamefont {P.}~\bibnamefont {Illien}}, \bibinfo {author}
  {\bibfnamefont {A.}~\bibnamefont {Law}}, \bibinfo {author} {\bibfnamefont
  {C.}~\bibnamefont {Mejía-Monasterio}}, \bibinfo {author} {\bibfnamefont
  {G.}~\bibnamefont {Oshanin}},\ and\ \bibinfo {author} {\bibfnamefont
  {R.}~\bibnamefont {Voituriez}},\ }\bibfield  {title} {\enquote {\bibinfo
  {title} {Geometry-{Induced} {Superdiffusion} in {Driven} {Crowded}
  {Systems}},}\ }\href {https://doi.org/10.1103/PhysRevLett.111.260601}
  {\bibfield  {journal} {\bibinfo  {journal} {Phys. Rev. Lett.}\ }\textbf
  {\bibinfo {volume} {111}},\ \bibinfo {pages} {260601} (\bibinfo {year}
  {2013})}\BibitemShut {NoStop}%
\bibitem [{\citenamefont {Wilson}\ and\ \citenamefont
  {Poon}(2011)}]{wilson_small-world_2011}%
  \BibitemOpen
  \bibfield  {author} {\bibinfo {author} {\bibfnamefont {L.~G.}\ \bibnamefont
  {Wilson}}\ and\ \bibinfo {author} {\bibfnamefont {W.~C.~K.}\ \bibnamefont
  {Poon}},\ }\bibfield  {title} {\enquote {\bibinfo {title} {Small-world
  rheology: an introduction to probe-based active microrheology},}\ }\href
  {https://doi.org/10.1039/C0CP01564D} {\bibfield  {journal} {\bibinfo
  {journal} {Phys. Chem. Chem. Phys.}\ }\textbf {\bibinfo {volume} {13}},\
  \bibinfo {pages} {10617--10630} (\bibinfo {year} {2011})}\BibitemShut
  {NoStop}%
\bibitem [{\citenamefont {Démery}\ and\ \citenamefont
  {Fodor}(2019)}]{demery_driven_2019}%
  \BibitemOpen
  \bibfield  {author} {\bibinfo {author} {\bibfnamefont {V.}~\bibnamefont
  {Démery}}\ and\ \bibinfo {author} {\bibfnamefont {E.}~\bibnamefont
  {Fodor}},\ }\bibfield  {title} {\enquote {\bibinfo {title} {Driven probe
  under harmonic confinement in a colloidal bath},}\ }\href
  {https://doi.org/10.1088/1742-5468/ab02e9} {\bibfield  {journal} {\bibinfo
  {journal} {J. Stat. Mech.}\ }\textbf {\bibinfo {volume} {2019}},\ \bibinfo
  {pages} {033202} (\bibinfo {year} {2019})}\BibitemShut {NoStop}%
\bibitem [{\citenamefont {Cugliandolo}(2011)}]{cugliandolo_effective_2011}%
  \BibitemOpen
  \bibfield  {author} {\bibinfo {author} {\bibfnamefont {L.~F.}\ \bibnamefont
  {Cugliandolo}},\ }\bibfield  {title} {\enquote {\bibinfo {title} {The
  effective temperature},}\ }\href
  {https://doi.org/10.1088/1751-8113/44/48/483001} {\bibfield  {journal}
  {\bibinfo  {journal} {J. Phys. A: Math. Theor.}\ }\textbf {\bibinfo {volume}
  {44}},\ \bibinfo {pages} {483001} (\bibinfo {year} {2011})}\BibitemShut
  {NoStop}%
\bibitem [{\citenamefont {Puglisi}, \citenamefont {Sarracino},\ and\
  \citenamefont {Vulpiani}(2017)}]{puglisi_temperature_2017}%
  \BibitemOpen
  \bibfield  {author} {\bibinfo {author} {\bibfnamefont {A.}~\bibnamefont
  {Puglisi}}, \bibinfo {author} {\bibfnamefont {A.}~\bibnamefont {Sarracino}},\
  and\ \bibinfo {author} {\bibfnamefont {A.}~\bibnamefont {Vulpiani}},\
  }\bibfield  {title} {\enquote {\bibinfo {title} {Temperature in and out of
  equilibrium: {A} review of concepts, tools and attempts},}\ }\href
  {https://doi.org/10.1016/j.physrep.2017.09.001} {\bibfield  {journal}
  {\bibinfo  {journal} {Physics Reports}\ }\textbf {\bibinfo {volume}
  {709-710}},\ \bibinfo {pages} {1--60} (\bibinfo {year} {2017})}\BibitemShut
  {NoStop}%
\bibitem [{\citenamefont {Brader}, \citenamefont {Cates},\ and\ \citenamefont
  {Fuchs}(2008)}]{brader_first-principles_2008}%
  \BibitemOpen
  \bibfield  {author} {\bibinfo {author} {\bibfnamefont {J.~M.}\ \bibnamefont
  {Brader}}, \bibinfo {author} {\bibfnamefont {M.~E.}\ \bibnamefont {Cates}},\
  and\ \bibinfo {author} {\bibfnamefont {M.}~\bibnamefont {Fuchs}},\ }\bibfield
   {title} {\enquote {\bibinfo {title} {First-{Principles} {Constitutive}
  {Equation} for {Suspension} {Rheology}},}\ }\href
  {https://doi.org/10.1103/PhysRevLett.101.138301} {\bibfield  {journal}
  {\bibinfo  {journal} {Phys. Rev. Lett.}\ }\textbf {\bibinfo {volume} {101}},\
  \bibinfo {pages} {138301} (\bibinfo {year} {2008})}\BibitemShut {NoStop}%
\bibitem [{\citenamefont {Krüger}, \citenamefont {Weysser},\ and\
  \citenamefont {Fuchs}(2011)}]{kruger_tagged-particle_2011}%
  \BibitemOpen
  \bibfield  {author} {\bibinfo {author} {\bibfnamefont {M.}~\bibnamefont
  {Krüger}}, \bibinfo {author} {\bibfnamefont {F.}~\bibnamefont {Weysser}},\
  and\ \bibinfo {author} {\bibfnamefont {M.}~\bibnamefont {Fuchs}},\ }\bibfield
   {title} {\enquote {\bibinfo {title} {Tagged-particle motion in glassy
  systems under shear: {Comparison} of mode coupling theory and {Brownian}
  dynamics simulations},}\ }\href {https://doi.org/10.1140/epje/i2011-11088-5}
  {\bibfield  {journal} {\bibinfo  {journal} {Eur. Phys. J. E}\ }\textbf
  {\bibinfo {volume} {34}},\ \bibinfo {pages} {88} (\bibinfo {year}
  {2011})}\BibitemShut {NoStop}%
\bibitem [{\citenamefont {Gazuz}\ and\ \citenamefont
  {Fuchs}(2013)}]{gazuz_nonlinear_2013}%
  \BibitemOpen
  \bibfield  {author} {\bibinfo {author} {\bibfnamefont {I.}~\bibnamefont
  {Gazuz}}\ and\ \bibinfo {author} {\bibfnamefont {M.}~\bibnamefont {Fuchs}},\
  }\bibfield  {title} {\enquote {\bibinfo {title} {Nonlinear microrheology of
  dense colloidal suspensions: {A} mode-coupling theory},}\ }\href
  {https://doi.org/10.1103/PhysRevE.87.032304} {\bibfield  {journal} {\bibinfo
  {journal} {Phys. Rev. E}\ }\textbf {\bibinfo {volume} {87}},\ \bibinfo
  {pages} {032304} (\bibinfo {year} {2013})}\BibitemShut {NoStop}%
\bibitem [{\citenamefont {Gruber}\ \emph {et~al.}(2016)\citenamefont {Gruber},
  \citenamefont {Abade}, \citenamefont {Puertas},\ and\ \citenamefont
  {Fuchs}}]{gruber_active_2016}%
  \BibitemOpen
  \bibfield  {author} {\bibinfo {author} {\bibfnamefont {M.}~\bibnamefont
  {Gruber}}, \bibinfo {author} {\bibfnamefont {G.~C.}\ \bibnamefont {Abade}},
  \bibinfo {author} {\bibfnamefont {A.~M.}\ \bibnamefont {Puertas}},\ and\
  \bibinfo {author} {\bibfnamefont {M.}~\bibnamefont {Fuchs}},\ }\bibfield
  {title} {\enquote {\bibinfo {title} {Active microrheology in a colloidal
  glass},}\ }\href {https://doi.org/10.1103/PhysRevE.94.042602} {\bibfield
  {journal} {\bibinfo  {journal} {Phys. Rev. E}\ }\textbf {\bibinfo {volume}
  {94}},\ \bibinfo {pages} {042602} (\bibinfo {year} {2016})}\BibitemShut
  {NoStop}%
\bibitem [{\citenamefont {Meyer}, \citenamefont {Voigtmann},\ and\
  \citenamefont {Schilling}(2017)}]{meyer_non-stationary_2017}%
  \BibitemOpen
  \bibfield  {author} {\bibinfo {author} {\bibfnamefont {H.}~\bibnamefont
  {Meyer}}, \bibinfo {author} {\bibfnamefont {T.}~\bibnamefont {Voigtmann}},\
  and\ \bibinfo {author} {\bibfnamefont {T.}~\bibnamefont {Schilling}},\
  }\bibfield  {title} {\enquote {\bibinfo {title} {On the non-stationary
  generalized {Langevin} equation},}\ }\href
  {https://doi.org/10.1063/1.5006980} {\bibfield  {journal} {\bibinfo
  {journal} {The Journal of Chemical Physics}\ }\textbf {\bibinfo {volume}
  {147}},\ \bibinfo {pages} {214110} (\bibinfo {year} {2017})}\BibitemShut
  {NoStop}%
\bibitem [{\citenamefont {te~Vrugt}\ and\ \citenamefont
  {Wittkowski}(2019)}]{te_vrugt_mori-zwanzig_2019}%
  \BibitemOpen
  \bibfield  {author} {\bibinfo {author} {\bibfnamefont {M.}~\bibnamefont
  {te~Vrugt}}\ and\ \bibinfo {author} {\bibfnamefont {R.}~\bibnamefont
  {Wittkowski}},\ }\bibfield  {title} {\enquote {\bibinfo {title}
  {Mori-{Zwanzig} projection operator formalism for far-from-equilibrium
  systems with time-dependent {Hamiltonians}},}\ }\href
  {https://doi.org/10.1103/PhysRevE.99.062118} {\bibfield  {journal} {\bibinfo
  {journal} {Phys. Rev. E}\ }\textbf {\bibinfo {volume} {99}},\ \bibinfo
  {pages} {062118} (\bibinfo {year} {2019})}\BibitemShut {NoStop}%
\bibitem [{\citenamefont {Meyer}, \citenamefont {Voigtmann},\ and\
  \citenamefont {Schilling}(2019)}]{meyer_dynamics_2019}%
  \BibitemOpen
  \bibfield  {author} {\bibinfo {author} {\bibfnamefont {H.}~\bibnamefont
  {Meyer}}, \bibinfo {author} {\bibfnamefont {T.}~\bibnamefont {Voigtmann}},\
  and\ \bibinfo {author} {\bibfnamefont {T.}~\bibnamefont {Schilling}},\
  }\bibfield  {title} {\enquote {\bibinfo {title} {On the dynamics of reaction
  coordinates in classical, time-dependent, many-body processes},}\ }\href
  {https://doi.org/10.1063/1.5090450} {\bibfield  {journal} {\bibinfo
  {journal} {The Journal of Chemical Physics}\ }\textbf {\bibinfo {volume}
  {150}},\ \bibinfo {pages} {174118} (\bibinfo {year} {2019})}\BibitemShut
  {NoStop}%
\bibitem [{\citenamefont {Glatzel}\ and\ \citenamefont
  {Schilling}(2021)}]{glatzel_interplay_2021}%
  \BibitemOpen
  \bibfield  {author} {\bibinfo {author} {\bibfnamefont {F.}~\bibnamefont
  {Glatzel}}\ and\ \bibinfo {author} {\bibfnamefont {T.}~\bibnamefont
  {Schilling}},\ }\bibfield  {title} {\enquote {\bibinfo {title} {The interplay
  between memory and potentials of mean force: {A} discussion on the structure
  of equations of motion for coarse-grained observables},}\ }\href
  {https://doi.org/10.1209/0295-5075/ac35ba} {\bibfield  {journal} {\bibinfo
  {journal} {EPL}\ }\textbf {\bibinfo {volume} {136}},\ \bibinfo {pages}
  {36001} (\bibinfo {year} {2021})}\BibitemShut {NoStop}%
\bibitem [{\citenamefont {Vroylandt}\ and\ \citenamefont
  {Monmarché}(2022)}]{vroylandt_position-dependent_2022}%
  \BibitemOpen
  \bibfield  {author} {\bibinfo {author} {\bibfnamefont {H.}~\bibnamefont
  {Vroylandt}}\ and\ \bibinfo {author} {\bibfnamefont {P.}~\bibnamefont
  {Monmarché}},\ }\bibfield  {title} {\enquote {\bibinfo {title}
  {Position-dependent memory kernel in generalized {Langevin} equations: theory
  and numerical estimation},}\ }\href {https://doi.org/10.1063/5.0094566}
  {\bibfield  {journal} {\bibinfo  {journal} {J. Chem. Phys.}\ }\textbf
  {\bibinfo {volume} {156}},\ \bibinfo {pages} {244105} (\bibinfo {year}
  {2022})}\BibitemShut {NoStop}%
\bibitem [{\citenamefont {Schilling}(2022)}]{schilling_coarse-grained_2022}%
  \BibitemOpen
  \bibfield  {author} {\bibinfo {author} {\bibfnamefont {T.}~\bibnamefont
  {Schilling}},\ }\bibfield  {title} {\enquote {\bibinfo {title}
  {Coarse-grained modelling out of equilibrium},}\ }\href
  {https://doi.org/10.1016/j.physrep.2022.04.006} {\bibfield  {journal}
  {\bibinfo  {journal} {Physics Reports}\ }\bibinfo {series} {Coarse-{Grained}
  {Modelling} {Out} of {Equilibrium}},\ \textbf {\bibinfo {volume} {972}},\
  \bibinfo {pages} {1--45} (\bibinfo {year} {2022})}\BibitemShut {NoStop}%
\bibitem [{\citenamefont {Netz}(2023)}]{netz_derivation_2023}%
  \BibitemOpen
  \bibfield  {author} {\bibinfo {author} {\bibfnamefont {R.~R.}\ \bibnamefont
  {Netz}},\ }\href {http://arxiv.org/abs/2310.00748} {\enquote {\bibinfo
  {title} {Derivation of the non-equilibrium generalized {Langevin} equation
  from a generic time-dependent {Hamiltonian}},}\ } (\bibinfo {year} {2023}),\
  \bibinfo {note} {arXiv:2310.00748 [cond-mat]}\BibitemShut {NoStop}%
\bibitem [{\citenamefont {Penna}, \citenamefont {Dzubiella},\ and\
  \citenamefont {Tarazona}(2003)}]{penna_dynamic_2003}%
  \BibitemOpen
  \bibfield  {author} {\bibinfo {author} {\bibfnamefont {F.}~\bibnamefont
  {Penna}}, \bibinfo {author} {\bibfnamefont {J.}~\bibnamefont {Dzubiella}},\
  and\ \bibinfo {author} {\bibfnamefont {P.}~\bibnamefont {Tarazona}},\
  }\bibfield  {title} {\enquote {\bibinfo {title} {Dynamic density functional
  study of a driven colloidal particle in polymer solutions},}\ }\href
  {https://doi.org/10.1103/PhysRevE.68.061407} {\bibfield  {journal} {\bibinfo
  {journal} {Phys. Rev. E}\ }\textbf {\bibinfo {volume} {68}},\ \bibinfo
  {pages} {061407} (\bibinfo {year} {2003})}\BibitemShut {NoStop}%
\bibitem [{\citenamefont {Rauscher}\ \emph {et~al.}(2007)\citenamefont
  {Rauscher}, \citenamefont {Domínguez}, \citenamefont {Krüger},\ and\
  \citenamefont {Penna}}]{rauscher_dynamic_2007}%
  \BibitemOpen
  \bibfield  {author} {\bibinfo {author} {\bibfnamefont {M.}~\bibnamefont
  {Rauscher}}, \bibinfo {author} {\bibfnamefont {A.}~\bibnamefont
  {Domínguez}}, \bibinfo {author} {\bibfnamefont {M.}~\bibnamefont
  {Krüger}},\ and\ \bibinfo {author} {\bibfnamefont {F.}~\bibnamefont
  {Penna}},\ }\bibfield  {title} {\enquote {\bibinfo {title} {A dynamic density
  functional theory for particles in a flowing solvent},}\ }\href
  {https://doi.org/10.1063/1.2806094} {\bibfield  {journal} {\bibinfo
  {journal} {The Journal of Chemical Physics}\ }\textbf {\bibinfo {volume}
  {127}},\ \bibinfo {pages} {244906} (\bibinfo {year} {2007})}\BibitemShut
  {NoStop}%
\bibitem [{\citenamefont {Gutsche}\ \emph {et~al.}(2008)\citenamefont
  {Gutsche}, \citenamefont {Kremer}, \citenamefont {Krüger}, \citenamefont
  {Rauscher}, \citenamefont {Weeber},\ and\ \citenamefont
  {Harting}}]{gutsche_colloids_2008}%
  \BibitemOpen
  \bibfield  {author} {\bibinfo {author} {\bibfnamefont {C.}~\bibnamefont
  {Gutsche}}, \bibinfo {author} {\bibfnamefont {F.}~\bibnamefont {Kremer}},
  \bibinfo {author} {\bibfnamefont {M.}~\bibnamefont {Krüger}}, \bibinfo
  {author} {\bibfnamefont {M.}~\bibnamefont {Rauscher}}, \bibinfo {author}
  {\bibfnamefont {R.}~\bibnamefont {Weeber}},\ and\ \bibinfo {author}
  {\bibfnamefont {J.}~\bibnamefont {Harting}},\ }\bibfield  {title} {\enquote
  {\bibinfo {title} {Colloids dragged through a polymer solution: {Experiment},
  theory, and simulation},}\ }\href {https://doi.org/10.1063/1.2965127}
  {\bibfield  {journal} {\bibinfo  {journal} {The Journal of Chemical Physics}\
  }\textbf {\bibinfo {volume} {129}},\ \bibinfo {pages} {084902} (\bibinfo
  {year} {2008})}\BibitemShut {NoStop}%
\bibitem [{\citenamefont {De~Las~Heras}\ and\ \citenamefont
  {Schmidt}(2018)}]{de_las_heras_velocity_2018}%
  \BibitemOpen
  \bibfield  {author} {\bibinfo {author} {\bibfnamefont {D.}~\bibnamefont
  {De~Las~Heras}}\ and\ \bibinfo {author} {\bibfnamefont {M.}~\bibnamefont
  {Schmidt}},\ }\bibfield  {title} {\enquote {\bibinfo {title} {Velocity
  {Gradient} {Power} {Functional} for {Brownian} {Dynamics}},}\ }\href
  {https://doi.org/10.1103/PhysRevLett.120.028001} {\bibfield  {journal}
  {\bibinfo  {journal} {Phys. Rev. Lett.}\ }\textbf {\bibinfo {volume} {120}},\
  \bibinfo {pages} {028001} (\bibinfo {year} {2018})}\BibitemShut {NoStop}%
\bibitem [{\citenamefont {De~Las~Heras}\ and\ \citenamefont
  {Schmidt}(2020)}]{de_las_heras_flow_2020}%
  \BibitemOpen
  \bibfield  {author} {\bibinfo {author} {\bibfnamefont {D.}~\bibnamefont
  {De~Las~Heras}}\ and\ \bibinfo {author} {\bibfnamefont {M.}~\bibnamefont
  {Schmidt}},\ }\bibfield  {title} {\enquote {\bibinfo {title} {Flow and
  {Structure} in {Nonequilibrium} {Brownian} {Many}-{Body} {Systems}},}\ }\href
  {https://doi.org/10.1103/PhysRevLett.125.018001} {\bibfield  {journal}
  {\bibinfo  {journal} {Phys. Rev. Lett.}\ }\textbf {\bibinfo {volume} {125}},\
  \bibinfo {pages} {018001} (\bibinfo {year} {2020})}\BibitemShut {NoStop}%
\bibitem [{\citenamefont {Schmidt}(2022)}]{schmidt_power_2022}%
  \BibitemOpen
  \bibfield  {author} {\bibinfo {author} {\bibfnamefont {M.}~\bibnamefont
  {Schmidt}},\ }\bibfield  {title} {\enquote {\bibinfo {title} {Power
  functional theory for many-body dynamics},}\ }\href
  {https://doi.org/10.1103/RevModPhys.94.015007} {\bibfield  {journal}
  {\bibinfo  {journal} {Rev. Mod. Phys.}\ }\textbf {\bibinfo {volume} {94}},\
  \bibinfo {pages} {015007} (\bibinfo {year} {2022})}\BibitemShut {NoStop}%
\bibitem [{\citenamefont {Asheichyk}, \citenamefont {Fuchs},\ and\
  \citenamefont {Krüger}(2021)}]{asheichyk_brownian_2021}%
  \BibitemOpen
  \bibfield  {author} {\bibinfo {author} {\bibfnamefont {K.}~\bibnamefont
  {Asheichyk}}, \bibinfo {author} {\bibfnamefont {M.}~\bibnamefont {Fuchs}},\
  and\ \bibinfo {author} {\bibfnamefont {M.}~\bibnamefont {Krüger}},\
  }\bibfield  {title} {\enquote {\bibinfo {title} {Brownian systems perturbed
  by mild shear: comparing response relations},}\ }\href
  {https://doi.org/10.1088/1361-648X/ac0c3c} {\bibfield  {journal} {\bibinfo
  {journal} {J. Phys.: Condens. Matter}\ }\textbf {\bibinfo {volume} {33}},\
  \bibinfo {pages} {405101} (\bibinfo {year} {2021})}\BibitemShut {NoStop}%
\bibitem [{\citenamefont {Krüger}\ and\ \citenamefont
  {Maes}(2016)}]{kruger_modified_2016}%
  \BibitemOpen
  \bibfield  {author} {\bibinfo {author} {\bibfnamefont {M.}~\bibnamefont
  {Krüger}}\ and\ \bibinfo {author} {\bibfnamefont {C.}~\bibnamefont {Maes}},\
  }\bibfield  {title} {\enquote {\bibinfo {title} {The modified {Langevin}
  description for probes in a nonlinear medium},}\ }\href
  {https://doi.org/10.1088/1361-648X/29/6/064004} {\bibfield  {journal}
  {\bibinfo  {journal} {J. Phys.: Condens. Matter}\ }\textbf {\bibinfo {volume}
  {29}},\ \bibinfo {pages} {064004} (\bibinfo {year} {2016})}\BibitemShut
  {NoStop}%
\bibitem [{\citenamefont {Müller}(2020)}]{muller_brownian_2020}%
  \BibitemOpen
  \bibfield  {author} {\bibinfo {author} {\bibfnamefont {B.}~\bibnamefont
  {Müller}},\ }\emph {\bibinfo {title} {Brownian {Particles} in
  {Nonequilibrium} {Solvents}}},\ \href {https://doi.org/10.53846/goediss-7798}
  {Ph.D. thesis},\ \bibinfo  {school} {Georg-{A}ugust-{U}niversit{\"a}t
  {G}{\"o}ttingen} (\bibinfo {year} {2020})\BibitemShut {NoStop}%
\bibitem [{\citenamefont {Colangeli}, \citenamefont {Maes},\ and\ \citenamefont
  {Wynants}(2011)}]{colangeli_meaningful_2011}%
  \BibitemOpen
  \bibfield  {author} {\bibinfo {author} {\bibfnamefont {M.}~\bibnamefont
  {Colangeli}}, \bibinfo {author} {\bibfnamefont {C.}~\bibnamefont {Maes}},\
  and\ \bibinfo {author} {\bibfnamefont {B.}~\bibnamefont {Wynants}},\
  }\bibfield  {title} {\enquote {\bibinfo {title} {A meaningful expansion
  around detailed balance},}\ }\href
  {https://doi.org/10.1088/1751-8113/44/9/095001} {\bibfield  {journal}
  {\bibinfo  {journal} {J. Phys. A: Math. Theor.}\ }\textbf {\bibinfo {volume}
  {44}},\ \bibinfo {pages} {095001} (\bibinfo {year} {2011})}\BibitemShut
  {NoStop}%
\bibitem [{\citenamefont {Basu}\ \emph {et~al.}(2015)\citenamefont {Basu},
  \citenamefont {Krüger}, \citenamefont {Lazarescu},\ and\ \citenamefont
  {Maes}}]{basu_frenetic_2015}%
  \BibitemOpen
  \bibfield  {author} {\bibinfo {author} {\bibfnamefont {U.}~\bibnamefont
  {Basu}}, \bibinfo {author} {\bibfnamefont {M.}~\bibnamefont {Krüger}},
  \bibinfo {author} {\bibfnamefont {A.}~\bibnamefont {Lazarescu}},\ and\
  \bibinfo {author} {\bibfnamefont {C.}~\bibnamefont {Maes}},\ }\bibfield
  {title} {\enquote {\bibinfo {title} {Frenetic aspects of second order
  response},}\ }\href {https://doi.org/10.1039/C4CP04977B} {\bibfield
  {journal} {\bibinfo  {journal} {Phys. Chem. Chem. Phys.}\ }\textbf {\bibinfo
  {volume} {17}},\ \bibinfo {pages} {6653--6666} (\bibinfo {year}
  {2015})}\BibitemShut {NoStop}%
\bibitem [{\citenamefont {Maes}(2020{\natexlab{a}})}]{maes_response_2020}%
  \BibitemOpen
  \bibfield  {author} {\bibinfo {author} {\bibfnamefont {C.}~\bibnamefont
  {Maes}},\ }\bibfield  {title} {\enquote {\bibinfo {title} {Response {Theory}:
  {A} {Trajectory}-{Based} {Approach}},}\ }\href
  {https://www.frontiersin.org/articles/10.3389/fphy.2020.00229} {\bibfield
  {journal} {\bibinfo  {journal} {Frontiers in Physics}\ }\textbf {\bibinfo
  {volume} {8}} (\bibinfo {year} {2020}{\natexlab{a}})}\BibitemShut {NoStop}%
\bibitem [{\citenamefont {Müller}\ \emph {et~al.}(2020)\citenamefont
  {Müller}, \citenamefont {Berner}, \citenamefont {Bechinger},\ and\
  \citenamefont {Krüger}}]{muller_properties_2020}%
  \BibitemOpen
  \bibfield  {author} {\bibinfo {author} {\bibfnamefont {B.}~\bibnamefont
  {Müller}}, \bibinfo {author} {\bibfnamefont {J.}~\bibnamefont {Berner}},
  \bibinfo {author} {\bibfnamefont {C.}~\bibnamefont {Bechinger}},\ and\
  \bibinfo {author} {\bibfnamefont {M.}~\bibnamefont {Krüger}},\ }\bibfield
  {title} {\enquote {\bibinfo {title} {Properties of a nonlinear bath:
  experiments, theory, and a stochastic {Prandtl}–{Tomlinson} model},}\
  }\href {https://doi.org/10.1088/1367-2630/ab6a39} {\bibfield  {journal}
  {\bibinfo  {journal} {New J. Phys.}\ }\textbf {\bibinfo {volume} {22}},\
  \bibinfo {pages} {023014} (\bibinfo {year} {2020})}\BibitemShut {NoStop}%
\bibitem [{\citenamefont {Jain}, \citenamefont {Ginot},\ and\ \citenamefont
  {Krüger}(2021)}]{jain_micro-rheology_2021}%
  \BibitemOpen
  \bibfield  {author} {\bibinfo {author} {\bibfnamefont {R.}~\bibnamefont
  {Jain}}, \bibinfo {author} {\bibfnamefont {F.}~\bibnamefont {Ginot}},\ and\
  \bibinfo {author} {\bibfnamefont {M.}~\bibnamefont {Krüger}},\ }\bibfield
  {title} {\enquote {\bibinfo {title} {Micro-rheology of a particle in a
  nonlinear bath: {Stochastic} {Prandtl}-{Tomlinson} model},}\ }\href
  {https://doi.org/10.1063/5.0062104} {\bibfield  {journal} {\bibinfo
  {journal} {Physics of Fluids}\ }\textbf {\bibinfo {volume} {33}},\ \bibinfo
  {pages} {103101} (\bibinfo {year} {2021})}\BibitemShut {NoStop}%
\bibitem [{Note1()}]{Note1}%
  \BibitemOpen
  \bibinfo {note} {The minus sign is included because it will yield the
  familiar sign convention for the case of indirect driving introduced
  below.}\BibitemShut {Stop}%
\bibitem [{CBp()}]{CBprivate}%
  \BibitemOpen
  \href@noop {} {}\bibinfo {note} {Clemens Bechinger, personal communication,
  2024}\BibitemShut {NoStop}%
\bibitem [{\citenamefont {Maes}(2020{\natexlab{b}})}]{maes_frenesy_2020}%
  \BibitemOpen
  \bibfield  {author} {\bibinfo {author} {\bibfnamefont {C.}~\bibnamefont
  {Maes}},\ }\bibfield  {title} {\enquote {\bibinfo {title} {Frenesy:
  {Time}-symmetric dynamical activity in nonequilibria},}\ }\href
  {https://doi.org/10.1016/j.physrep.2020.01.002} {\bibfield  {journal}
  {\bibinfo  {journal} {Physics Reports}\ }\bibinfo {series} {Frenesy:
  time-symmetric dynamical activity in nonequilibria},\ \textbf {\bibinfo
  {volume} {850}},\ \bibinfo {pages} {1--33} (\bibinfo {year}
  {2020}{\natexlab{b}})}\BibitemShut {NoStop}%
\bibitem [{\citenamefont {Onsager}\ and\ \citenamefont
  {Machlup}(1953)}]{onsager_fluctuations_1953}%
  \BibitemOpen
  \bibfield  {author} {\bibinfo {author} {\bibfnamefont {L.}~\bibnamefont
  {Onsager}}\ and\ \bibinfo {author} {\bibfnamefont {S.}~\bibnamefont
  {Machlup}},\ }\bibfield  {title} {\enquote {\bibinfo {title} {Fluctuations
  and {Irreversible} {Processes}},}\ }\href
  {https://doi.org/10.1103/PhysRev.91.1505} {\bibfield  {journal} {\bibinfo
  {journal} {Phys. Rev.}\ }\textbf {\bibinfo {volume} {91}},\ \bibinfo {pages}
  {1505--1512} (\bibinfo {year} {1953})}\BibitemShut {NoStop}%
\bibitem [{\citenamefont {Baiesi}, \citenamefont {Maes},\ and\ \citenamefont
  {Wynants}(2009)}]{baiesi_fluctuations_2009}%
  \BibitemOpen
  \bibfield  {author} {\bibinfo {author} {\bibfnamefont {M.}~\bibnamefont
  {Baiesi}}, \bibinfo {author} {\bibfnamefont {C.}~\bibnamefont {Maes}},\ and\
  \bibinfo {author} {\bibfnamefont {B.}~\bibnamefont {Wynants}},\ }\bibfield
  {title} {\enquote {\bibinfo {title} {Fluctuations and {Response} of
  {Nonequilibrium} {States}},}\ }\href
  {https://doi.org/10.1103/PhysRevLett.103.010602} {\bibfield  {journal}
  {\bibinfo  {journal} {Phys. Rev. Lett.}\ }\textbf {\bibinfo {volume} {103}},\
  \bibinfo {pages} {010602} (\bibinfo {year} {2009})}\BibitemShut {NoStop}%
\bibitem [{\citenamefont {Maes}\ and\ \citenamefont
  {Netočný}(2003)}]{maes_time-reversal_2003}%
  \BibitemOpen
  \bibfield  {author} {\bibinfo {author} {\bibfnamefont {C.}~\bibnamefont
  {Maes}}\ and\ \bibinfo {author} {\bibfnamefont {K.}~\bibnamefont
  {Netočný}},\ }\bibfield  {title} {\enquote {\bibinfo {title}
  {Time-{Reversal} and {Entropy}},}\ }\href
  {https://doi.org/10.1023/A:1021026930129} {\bibfield  {journal} {\bibinfo
  {journal} {Journal of Statistical Physics}\ }\textbf {\bibinfo {volume}
  {110}},\ \bibinfo {pages} {269--310} (\bibinfo {year} {2003})}\BibitemShut
  {NoStop}%
\bibitem [{\citenamefont {Seifert}(2012)}]{seifert_stochastic_2012}%
  \BibitemOpen
  \bibfield  {author} {\bibinfo {author} {\bibfnamefont {U.}~\bibnamefont
  {Seifert}},\ }\bibfield  {title} {\enquote {\bibinfo {title} {Stochastic
  thermodynamics, fluctuation theorems and molecular machines},}\ }\href
  {https://doi.org/10.1088/0034-4885/75/12/126001} {\bibfield  {journal}
  {\bibinfo  {journal} {Rep. Prog. Phys.}\ }\textbf {\bibinfo {volume} {75}},\
  \bibinfo {pages} {126001} (\bibinfo {year} {2012})}\BibitemShut {NoStop}%
\bibitem [{\citenamefont {Holsten}\ and\ \citenamefont
  {Krüger}(2021)}]{holsten_thermodynamic_2021}%
  \BibitemOpen
  \bibfield  {author} {\bibinfo {author} {\bibfnamefont {T.}~\bibnamefont
  {Holsten}}\ and\ \bibinfo {author} {\bibfnamefont {M.}~\bibnamefont
  {Krüger}},\ }\bibfield  {title} {\enquote {\bibinfo {title} {Thermodynamic
  nonlinear response relation},}\ }\href
  {https://doi.org/10.1103/PhysRevE.103.032116} {\bibfield  {journal} {\bibinfo
   {journal} {Phys. Rev. E}\ }\textbf {\bibinfo {volume} {103}},\ \bibinfo
  {pages} {032116} (\bibinfo {year} {2021})}\BibitemShut {NoStop}%
\bibitem [{\citenamefont {Kardar}(2007)}]{kardar_statistical_2007}%
  \BibitemOpen
  \bibfield  {author} {\bibinfo {author} {\bibfnamefont {M.}~\bibnamefont
  {Kardar}},\ }\href {https://doi.org/10.1017/CBO9780511815881} {\emph
  {\bibinfo {title} {Statistical {Physics} of {Fields}}}}\ (\bibinfo
  {publisher} {Cambridge University Press},\ \bibinfo {address} {Cambridge},\
  \bibinfo {year} {2007})\BibitemShut {NoStop}%
\bibitem [{\citenamefont {Asheichyk}\ \emph {et~al.}(2019)\citenamefont
  {Asheichyk}, \citenamefont {Solon}, \citenamefont {Rohwer},\ and\
  \citenamefont {Krüger}}]{asheichyk_response_2019}%
  \BibitemOpen
  \bibfield  {author} {\bibinfo {author} {\bibfnamefont {K.}~\bibnamefont
  {Asheichyk}}, \bibinfo {author} {\bibfnamefont {A.~P.}\ \bibnamefont
  {Solon}}, \bibinfo {author} {\bibfnamefont {C.~M.}\ \bibnamefont {Rohwer}},\
  and\ \bibinfo {author} {\bibfnamefont {M.}~\bibnamefont {Krüger}},\
  }\bibfield  {title} {\enquote {\bibinfo {title} {Response of active
  {Brownian} particles to shear flow},}\ }\href
  {https://doi.org/10.1063/1.5086495} {\bibfield  {journal} {\bibinfo
  {journal} {J. Chem. Phys.}\ }\textbf {\bibinfo {volume} {150}},\ \bibinfo
  {pages} {144111} (\bibinfo {year} {2019})}\BibitemShut {NoStop}%
\bibitem [{\citenamefont {Peccati}\ and\ \citenamefont
  {Taqqu}(2011)}]{peccati_wiener_2011}%
  \BibitemOpen
  \bibfield  {author} {\bibinfo {author} {\bibfnamefont {G.}~\bibnamefont
  {Peccati}}\ and\ \bibinfo {author} {\bibfnamefont {M.~S.}\ \bibnamefont
  {Taqqu}},\ }\href {https://doi.org/10.1007/978-88-470-1679-8} {\emph
  {\bibinfo {title} {Wiener {Chaos}: {Moments}, {Cumulants} and {Diagrams}}}},\
  \bibinfo {series} {Bocconi \& {Springer} {Series}}, Vol.~\bibinfo {volume}
  {1}\ (\bibinfo  {publisher} {Springer Milan},\ \bibinfo {address} {Milano},\
  \bibinfo {year} {2011})\BibitemShut {NoStop}%
\bibitem [{\citenamefont {di~Bruno}(1855)}]{bruno_sullo_1855}%
  \BibitemOpen
  \bibfield  {author} {\bibinfo {author} {\bibfnamefont {F.}~\bibnamefont
  {di~Bruno}},\ }\bibfield  {title} {\enquote {\bibinfo {title} {Sullo sviluppo
  delle funzioni},}\ }\href@noop {} {\bibfield  {journal} {\bibinfo  {journal}
  {Ann. Sci. Mat. Fis., Roma}\ }\textbf {\bibinfo {volume} {6}},\ \bibinfo
  {pages} {479--480} (\bibinfo {year} {1855})}\BibitemShut {NoStop}%
\end{thebibliography}

%

\end{document}